\documentclass[12pt,letterpaper]{article}
\usepackage[a4paper, total={7in, 10in}]{geometry}

\usepackage{graphicx}
\usepackage{helvet}
\usepackage{authblk}
\usepackage{hyperref}
\usepackage{amsmath} 
\usepackage{amssymb} 
\usepackage{orcidlink} 
\usepackage[super,comma,sort&compress]  
   {natbib}

\usepackage{xurl}

\usepackage{xurl}
\usepackage{makecell}
\usepackage{multirow}
\usepackage{kantlipsum, lipsum}
\usepackage{pstricks, pst-node}
\usepackage{verbatim}
\usepackage{scalerel}
\usepackage{booktabs}
\usepackage{enumitem}
\usepackage{xspace}
\usepackage{bm}
\usepackage{bbm}
\usepackage{mathtools}
\usepackage{soul}
\usepackage{epsfig}
\usepackage{csquotes}
\usepackage{setspace}
\usepackage{colortbl}
\usepackage{threeparttable}
\usepackage{tabularx,ragged2e} 
\usepackage{placeins}
\usepackage{amsthm}
\newtheorem{theorem}{Theorem}
\newtheorem{assumption}{Assumption}

\usepackage{caption}
\makeatletter
\renewcommand{\maketitle}{\bgroup\setlength{\parindent}{0pt}
\begin{flushleft}
  \textbf{\@title}
  
  \@author
\end{flushleft}\egroup}
\makeatother

\title{\LARGE\bfseries Knowledge-enhanced Pretraining for Vision-\\language Pathology Foundation Model on Cancer Diagnosis \\[8pt]}
\date{}

\author[1]{Xiao Zhou}
\author[1,2]{Luoyi Sun}
\author[3]{Dexuan He}
\author[4]{Wenbin Guan}
\author[5]{Ge Wang}
\author[4]{Ruifen Wang}
\author[4]{Lifeng Wang}
\author[6]{Xiaojun Yuan}
\author[7]{Xin Sun}
\author[1,3,*]{ Ya Zhang}
\author[8,*]{Kun Sun}
\author[3,*]{Yanfeng Wang}
\author[1,3,9,*]{\\ Weidi Xie}

\affil[1]{\small Shanghai Artificial Intelligence Laboratory, Shanghai 200232, China}
\affil[2]{\small Zhejiang University, Zhejiang 310058, China}
\affil[3]{\small School of Artificial Intelligence, Shanghai Jiao Tong University, Shanghai 200230, China}
\affil[4]{\small Department of Pathology, Xinhua Hospital Affiliated to Shanghai Jiao Tong University School of Medicine, Shanghai 200092, China}
\affil[5]{\small Department of Oral Pathology, Shanghai Ninth People's Hospital, Shanghai Jiao Tong University School of Medicine, Shanghai 200011, China}
\affil[6]{\small Department of Pediatric Hematology/Oncology, Xinhua Hospital Affiliated to Shanghai Jiao Tong University School of Medicine, Shanghai 200092, China}
\affil[7]{\small Clinical Research and Innovation Unit, Xinhua Hospital Affiliated to Shanghai Jiao Tong University School of Medicine, Shanghai 200092, China}
\affil[8]{\small Department of Pediatric Cardiology, Xinhua Hospital Affiliated to Shanghai Jiao Tong University School of Medicine, Shanghai 200092, China}
\affil[9]{\small Lead contact}

\affil[*]{\small Correspondence: ya\_zhang@sjtu.edu.cn, sunkun@xinhuamed.com.cn, wangyanfeng@sjtu.edu.cn, weidi@sjtu.edu.cn}

\begin{document}

\maketitle

\section*{SUMMARY}

Vision–language foundation models have shown great promise in computational pathology but remain primarily data-driven, lacking explicit integration of medical knowledge. We introduce \textbf{KEEP} (\textbf{K}nowledg\textbf{E}-\textbf{E}nhanced \textbf{P}athology), a foundation model that systematically incorporates disease knowledge into pre-training for cancer diagnosis. KEEP leverages a comprehensive disease knowledge graph encompassing 11,454 diseases and 139,143 attributes to reorganize millions of pathology image–text pairs into 143,000 semantically structured groups aligned with disease ontology hierarchies. This knowledge-enhanced pre-training aligns visual and textual representations within hierarchical semantic spaces, enabling deeper understanding of disease relationships and morphological patterns. Across 18 public benchmarks (over 14,000 whole-slide images) and 4 institutional rare cancer datasets (926 cases), KEEP consistently outperformed existing foundation models, showing substantial gains for rare subtypes. These results establish knowledge-enhanced vision–language modeling as a powerful paradigm for advancing computational pathology.


\section*{KEYWORDS}

pathology foundation model, cancer diagnosis, vision-language, knowledge enhancement

\section*{INTRODUCTION}

Pathology diagnosis remains the golden standard in clinical applications for cancer diagnosis. Over the past decade, advancements in deep learning for computer vision have catalyzed significant progress in computational pathology, 
resulting in the development of specialized models based on both full supervision~\cite{shaban2020context,shao2021transmil,lin2023interventional,chan2023histopathology,huang2023conslide} or weak supervision~\cite{campanella2019clinical,zhou2020lirnet,lu2021data,chen2023rankmix,li2023task, wang2024pathology, el2024whole, neidlinger2024benchmarking}. While these approaches show promise, they are generally limited by the high cost and scarcity of annotations, as well as their restricted generalizability across diverse datasets. To address these limitations, self-supervised learning (SSL) strategies~\cite{wang2022transformer,chen2022fast,chen2022scaling,kang2023benchmarking} have emerged as a promising alternative, enabling to pre-train the model on large collections of unlabeled pathological images, acting as a versatile feature extractor for a series of downstream tasks~\cite{filiot2023scaling, chen2024uni, vorontsov2024foundation, xu2024whole, zimmermann2024virchow2}. However, the vision-only SSL models still require fine-tuning on diverse labeled datasets for specific tasks, limiting their scalability to low-annotation settings, particularly in rare cancer subtype classification tasks.

\begin{figure}
    \centering
    \includegraphics[scale = 0.75]{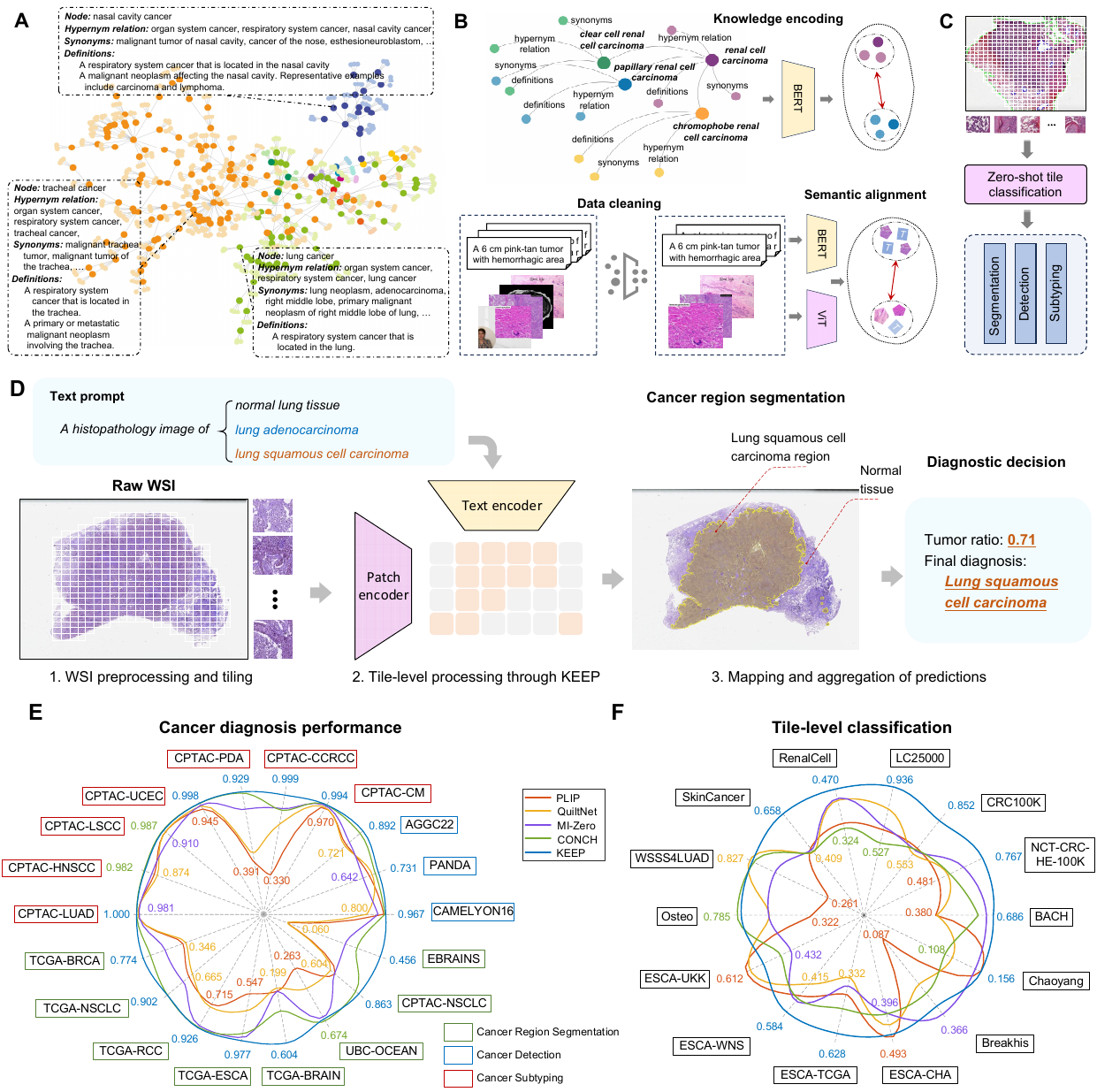}
    \caption{\textbf{Overview of KEEP.} 
    \textbf{A.} Example disease structure in the constructed knowledge graph. 
    Each node represents a disease, consisting of three attribute types: 
    hierarchical relations, synonyms, and definitions, as indicated by the dashed line box.
    \textbf{B.} The knowledge encoding and vision-language alignment stage for the KEEP model. A BERT-based text encoder is initially trained to encode the disease knowledge through metric learning. A knowledge-enhanced vision-language pre-training approach is proposed to align pathology semantic groups with filtered images and augmented captions.
    \textbf{C.} For downstream cancer diagnostic tasks, including cancer region segmentation, cancer detection, and cancer subtyping, whole slide images (WSIs) are divided into tile images for zero-shot classification, with the results of each tile combined to determine the final diagnostic decision. The text prompt for zero-shot classification is [template + disease name], for instance, \textit{A histopathology image of lung adenocarcinoma.}
    \textbf{D.} The flowchart of cancer diagnosis, including WSI pre-processing and titling, Tile-level processing through KEEP model, and mapping and aggregation of predictions.
    \textbf{E.} Performance comparison of cancer diagnosis with the state-of-the-art methods on 18 public benchmarks of more than 14,000 WSIs. The details of all datasets can be found in Table S1.
    \textbf{F.} Performance comparison of tile-level classification with the state-of-the-art methods on 14 benchmarks. The inner and outer numbers indicate the worst and best results, respectively.  Also see Figure~\ref{fig:supp_statictis} and Table S1.
    }
    \label{fig:main}
\end{figure}

Recently, the rise of vision-language models~\cite{radford2021learning,jia2021scaling} has enabled a new paradigm for computational pathology, offering novel avenues in cancer diagnosis. By jointly leveraging visual and textual data, vision-language models introduce free-text descriptions as supervision signals for pathology image representation learning, potentially improving diagnostic accuracy even in data-scarce settings. 
This approach could enhance generalizability and reduce reliance on extensive labeled datasets, addressing the limitations of vision-only models in distinguishing complex cancer subtypes. To create a joint embedding space for vision and language, 
existing models are trained on pathology image-text pairs gathered from in-house resources (MI-Zero~\cite{lu2023visual}, CONCH~\cite{lu2024visual}, and PRISM~\cite{shaikovski2024prism}) or public websites, such as Twitter (PLIP~\cite{huang2023visual}) and YouTube videos (QuiltNet~\cite{ikezogwo2024quilt}), employing straightforward contrastive learning to align images with their corresponding captions. 

Despite achieving impressive performance across various downstream tasks, existing pathology vision-language models, including PLIP and QuiltNet, face significant limitations due to the relatively small scale of pathology image-text datasets like OpenPath~\cite{huang2023visual} and Quilt1M~\cite{ikezogwo2024quilt}. 
Compared to the expansive datasets used in general computer vision, these pathology-specific resources are orders of magnitude smaller and often sourced from non-professional websites, leading to considerable data noise and limited quality, for example, the annotations accompanying these images tend to be brief, unstructured, and lacking in comprehensive medical knowledge. Such deficiencies hinder the models’ ability to accurately recognize and differentiate between various disease manifestations and their corresponding pathological features.

Zero-shot cancer diagnosis is a key downstream application of pathology vision-language foundation models\cite{lu2024visual,shaikovski2024prism}, that well suits the scenarios for diagnosing rare tumors with very few labeled cases. Modern foundation models, typically fed with small gridded tiles from whole-slide image (WSI), integrate embedding features (in vision-only models~\cite{chen2024uni}) or predicted labels (in vision-language models\cite{lu2024visual}) to derive final diagnostic decisions. While vision-language models offer a more explainable approach by explicitly identifying cancerous tiles, their performance in diagnosing rare diseases is still limited. 
Here, we introduce \textbf{KEEP} (\textbf{K}nowledg\textbf{E}-\textbf{E}nhanced \textbf{P}athology), a vision-language foundation model that integrates hierarchical medical knowledge with data-driven pretraining. To achieve this, we constructed a disease knowledge graph (KG) encompassing 11,454 disease entities from authoritative ontologies (Figure~\ref{fig:main}A) like the Disease Ontology (DO)~\cite{schriml2012disease} and the Unified Medical Language System (UMLS)~\cite{bodenreider2004unified}, which guides a pretraining framework that aligns filtered, semantically grouped pathology image-text pairs (Figure~\ref{fig:supp_statictis}) with clinical hierarchies (Figure~\ref{fig:main}B). We performed a comprehensive evaluation on the largest scale to date, spanning 18 public benchmarks ($>$14,000 WSIs) and 4 in-house rare cancer datasets (Table S1), covering critical tasks including cancer region segmentation, detection, and subtyping from WSIs (Figure~\ref{fig:main}C,D) and cross-modal retrieval from tile-level images. KEEP demonstrates superior performance over existing state-of-the-art models (Figure~\ref{fig:main}E,F): it achieves an average sensitivity of 0.898 (at 0.95 specificity) in tumor detection across 7 cancer types, significantly outperforming CHIEF~\cite{wang2024pathology}, and surpasses CONCH~\cite{lu2024visual} by 8.5 points in balanced accuracy for subtyping 30 rare brain cancers. Furthermore, ablation studies confirm that knowledge integration yields average improvements of 7.3\% and 7.2\% in segmentation and subtyping metrics, validating the critical role of structured medical knowledge in enhancing diagnostic precision for both common and rare cancer diagnosis.

\section*{RESULTS}

\subsection*{Overview of KEEP}

KEEP is a vision-language foundation model that leverages a disease knowledge graph to enhance both prediction performance and explainability in cancer diagnosis. By aligning semantic groups with a well-defined knowledge structure, KEEP outperforms existing models like PLIP and CONCH, which rely on naive contrastive learning of image-text pairs. This knowledge-driven approach deepens the model’s understanding of various disease characteristics and ensures stronger semantic alignment across diagnostic tasks. 

\begin{figure}
  \centering
  \includegraphics[scale=0.8]{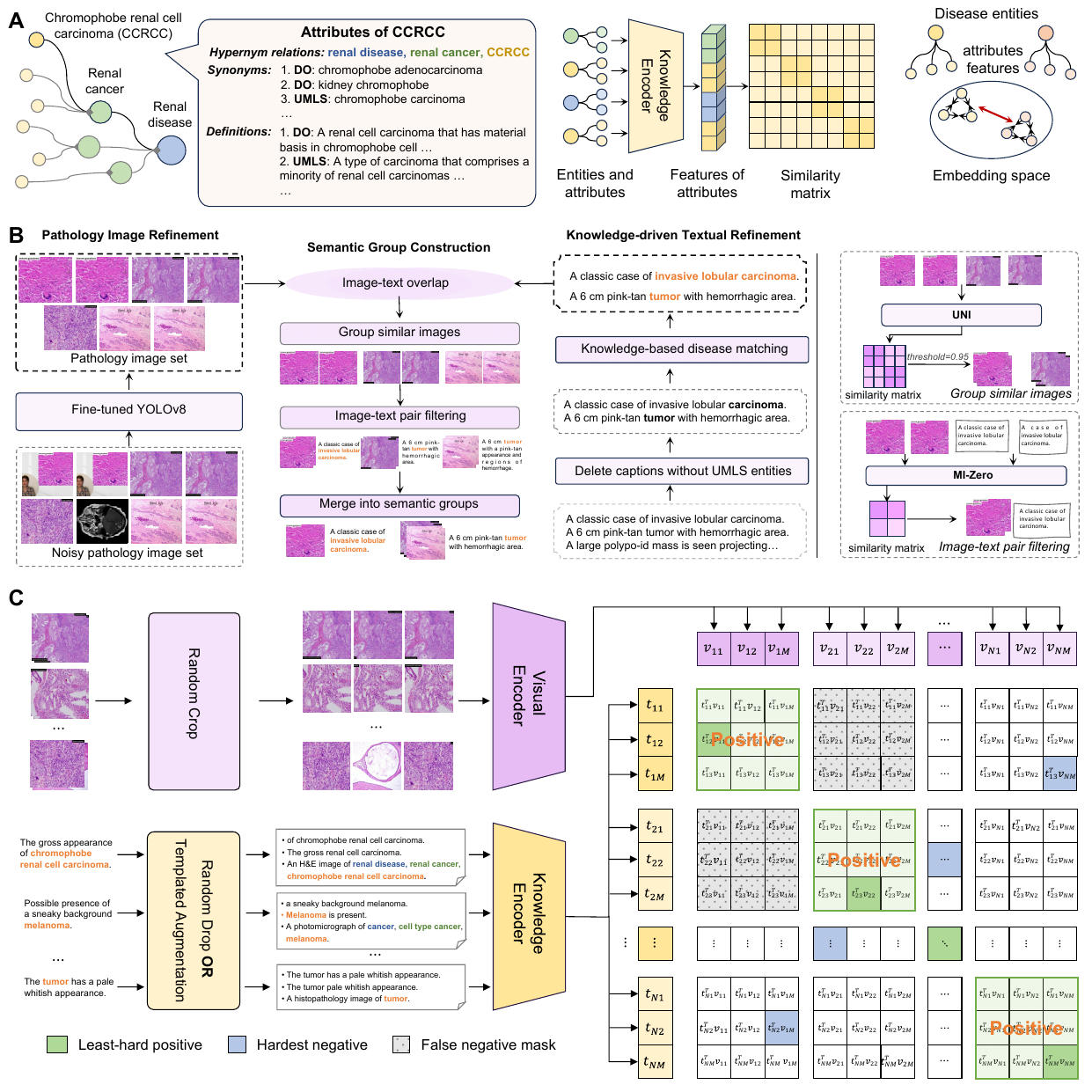}
  \caption{\textbf{Model architecture of KEEP}. 
  \textbf{A}. Disease knowledge encoding. We establish a knowledge graph that includes hypernym relations, synonyms, and definitions of diseases, and pretrain a disease knowledge encoder. Diseases at different levels are represented by different colors. 
  \textbf{B}. Knowledge-guided dataset structuring. We fine-tune YOLOv8 to remove noise in the pathology image dataset, extract medical entities from the captions, align the diseases in the captions with the diseases and synonyms in the knowledge graph, and cluster the filtered image and text data into semantic groups. The right side illustrates two specific methods used during the clustering process. 
  \textbf{C}. Knowledge-enhanced vision-language pretraining. We perform cropping and random dropping augmentations on the images and texts, and paraphrase captions that contain diseases using templates. During the training process, to mitigate the impact of false negatives, we design strategies for positive mining, hardest negative, and false negative elimination. Also see Figure~\ref{fig:supp_disease_chain} and Table S2.}
  \label{fig:methods}
\end{figure}

Specifically, we first curate a hierarchical knowledge graph~(KG) that consists of 11,454 human diseases and corresponding disease attributes, 
including disease synonyms, definitions, and hypernym relations (Figure~\ref{fig:methods}A). For instance, {\em lung squamous cell carcinoma, also known as epidermoid cell carcinoma of the lung }(\textbf{synonym}){\em, is a carcinoma that derives from squamous epithelial cells }(\textbf{definition}){\em, which is a subtype of non-small cell lung cancer and also a subtype of squamous cell carcinoma} (\textbf{hypernym relations}). 
We then train a language model to encode this knowledge graph, 
with the goal of learning the hierarchical relationships between different diseases (Figure~\ref{fig:methods}A and Figure~\ref{fig:supp_disease_chain}). 

Guided by the curated disease knowledge graph, 
we conduct thorough dataset cleaning and reorganise the noisy image-text data from OpenPath~\cite{huang2023visual} and Quilt1M~\cite{ikezogwo2024quilt} into 143k semantic groups tied by well-defined hypernym relations (Figure~\ref{fig:methods}B). 
More details on dataset cleaning can be found in Methods.
The statistics of semantic groups can be found in Figure~\ref{fig:supp_statictis}.
Subsequently, the knowledge encoder is used to guide the vision-language representation learning through a novel semantic-level alignment~(Figure~\ref{fig:methods}C). 
More details of the KEEP model can be found in Methods. 
We evaluate our model for cancer diagnostic tasks in a zero-shot manner, where the text prompt is [template + disease name], for instance, \textit{a histopathology image of lung adenocarcinoma}.
The results of zero-shot cancer region segmentation, cancer detection, cancer subtyping, and tile image profiling are exhibited in the following subsections.

For clinical application, as shown in Figure~\ref{fig:main}D, KEEP can be integrated into clinical workflows. Specifically, a whole‑slide image (WSI) is divided into fixed‑size tiles (e.g., 256 × 256 pixels at 20×) for efficient processing while preserving histological detail.
KEEP performs zero‑shot inference by matching tile embeddings with text prompts (e.g., “Lung adenocarcinoma”, “Normal tissue”) to assign labels.
Tile‑level predictions are aggregated into a diagnostic map, and the tumor ratio is used for slide‑level diagnosis.
The resulting map provides interpretable, quantitative guidance that supports and accelerates pathologists’ diagnostic decisions.

\subsection*{KEEP enhances slide-level cancer region segmentation}

\begin{figure}
  \centering
  \includegraphics[scale=0.8]{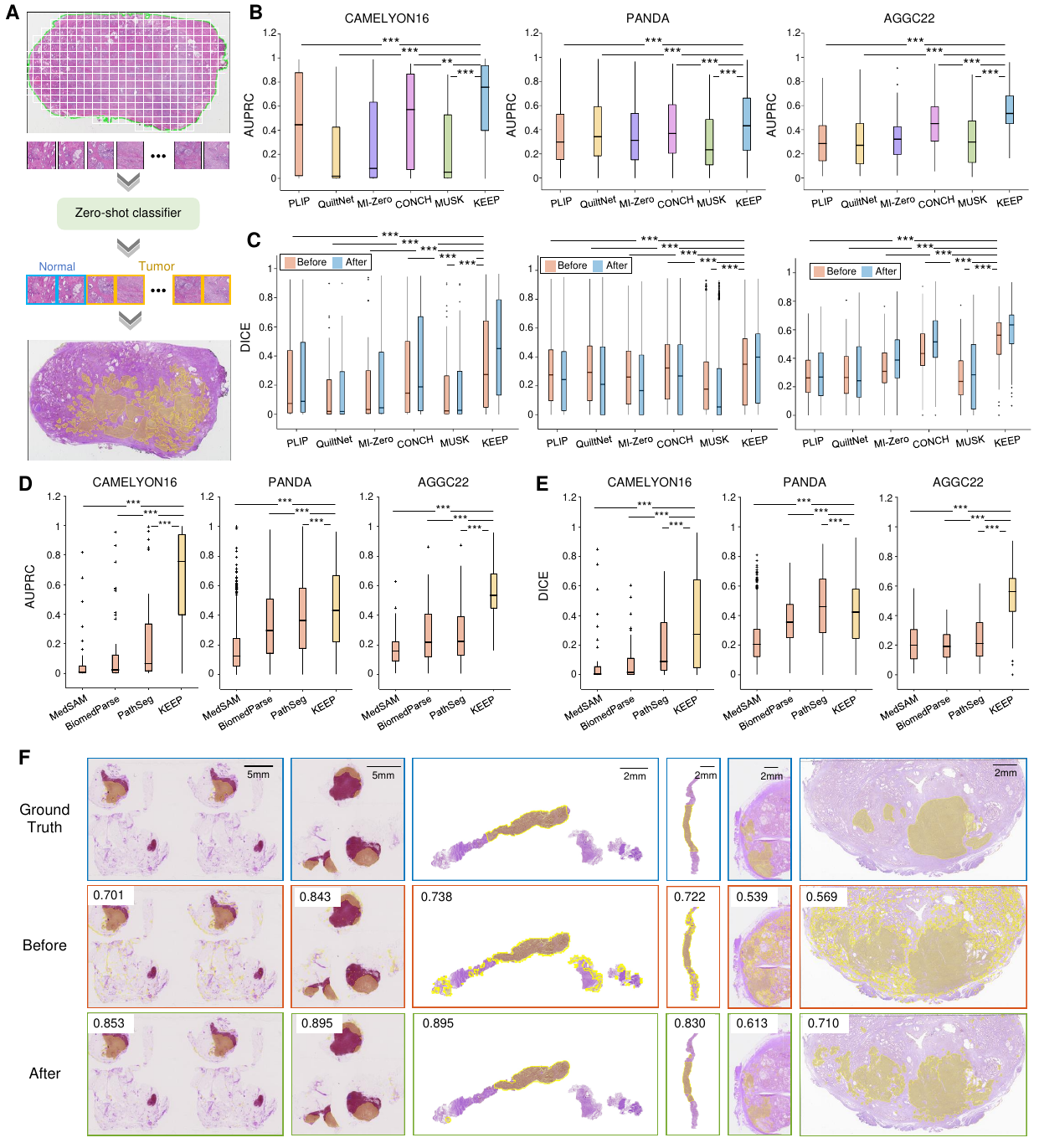}
  \caption{\textbf{KEEP enhances slide-level cancer region segmentation}. 
  \textbf{A}. The scheme of zero-shot segmentation on WSIs, where individual tiles undergo binary classification and are then combined to delineate the cancerous region. 
  \textbf{B-C}. Performance comparisons of AUPRC and DICE scores for various models, including PLIP~\cite{huang2023visual}, QuiltNet~\cite{ikezogwo2024quilt}, MI-Zero~\cite{lu2023visual}, CONCH~\cite{lu2024visual}, and MUSK~\cite{xiang2025vision}, and our proposed KEEP, across three WSI datasets: CAMELYON16~\cite{bejnordi2017camelyon16} (48 WSIs), PANDA~\cite{bulten2022panda} (10,494 WSIs), and AGGC22~\cite{huo2024aggc22} (128 WSIs). The DICE score is calculated using the average threshold corresponding to the optimal cutoff point of ROC curves in each dataset.  “Before” and “after” represent the segmentation results before and after post-processing with a morphological opening operation, which removes small noisy regions while preserving the shape of larger structures. The box plots present the median, first, and third quartiles of results. The paired t-test is used to assess the statistical significance between the performance distributions of different models. ** denotes $P < 0.01$, and *** denotes $P < 0.001$.
  \textbf{D-E}. Performance comparisons of AUPRC and DICE scores between KEEP and text-based segmentation models, including MedSAM~\cite{ma2024medsam}, BiomedParse~\cite{zhao2025biomedparse}, and PathSeg~\cite{chen2025pathseg}.
  The box plots present the median, first, and third quartiles of results, with $\mu$ indicating the average performance.  The paired t-test is used to assess the statistical significance between the performance distributions of different models. ** denotes $P < 0.01$, and *** denotes $P < 0.001$.
  \textbf{F}. Exemplary segmentation results from three datasets (the first two for CAMELYON16, the middle two for PANDA, and the last two for AGGC22) before and after post-processing. The number in the top-left of each result image suggests the DICE score. Also see Figure~\ref{fig:supp_seg} and Table S3.}
  \label{fig:result_seg}
\end{figure}

Segmenting cancerous regions from WSIs to define the region of interest (ROI) is critical for subsequent morphological profiling in cancer diagnosis. Traditional approaches rely on labor-intensive, 
task-specific manual annotations to train slide-level segmentation models, 
a process that is both costly and time-consuming, limiting the scalability of computational pathology. In contrast, vision-language foundation models can perform zero-shot segmentation of cancerous regions by classifying image tiles, enabling coarse-grained segmentation of WSIs (Figure~\ref{fig:result_seg}A).

Following the common practice~\cite{lu2024visual}, 
we divide the tissue regions of each WSI into small tiles ($224 \times 224$) with 75\% overlap, and average the classification scores in the overlapping areas to generate the final segmentation map. 
We compare the performance of KEEP with five other pathology vision-language models—PLIP~\cite{huang2023visual}, QuiltNet~\cite{ikezogwo2024quilt}, MI-Zero~\cite{lu2023visual}, CONCH~\cite{lu2024visual}, and MUSK~\cite{xiang2025vision}—on three datasets: CAMELYON16~\cite{bejnordi2017camelyon16} (48 WSIs), PANDA~\cite{bulten2022panda} (10,494 WSIs), and AGGC22~\cite{huo2024aggc22} (128 WSIs). 
We adopt the same approach as CONCH~\cite{lu2024visual}, 
ensembling 50 text prompts for each experiment and using the softmax function to normalize the similarities between the tile image and binary text prompts.
Segmentation performance is evaluated using the area under the curve (AUROC and AUPRC), DICE score, and Average Symmetric Surface Distance (ASSD) across all WSIs, as shown in Figure~\ref{fig:result_seg}B, Figure~\ref{fig:supp_seg}A-C,
Figure~\ref{fig:result_seg}C, and 
Table S3,
respectively.
KEEP consistently outperforms the other models across all datasets. 
Notably, it achieves an average DICE score improvement of 6.8 and 8.1 points over the state-of-the-art model CONCH on the CAMELYON16 and AGGC22 datasets, respectively.

Since each tile in WSIs is independently classified as a cancerous or normal patch, 
the segmentation results prone to exhibit scattered false positives. 
We employ a simple post-processing with morphological operations. 
Specifically, we treat the prediction for each tile as a single pixel and apply a morphological opening operation~({\em i.e.}, erosion followed by dilation) to filter out small, isolated noisy regions. The post-processed results~(noted by After), demonstrate a substantial improvement over KEEP in all datasets (9 points for CAMELYON16~\cite{bejnordi2017camelyon16}, 3 points for PANDA~\cite{bulten2022panda}, and 6.7 points for AGGC22~\cite{huo2024aggc22}) (Figure~\ref{fig:result_seg}C). Example segmentation results of KEEP before and after post-processing are visualized in Figure~\ref{fig:result_seg}F, with examples from CAMELYON16, PANDA, and AGGC22 in the left, middle, and right two columns, respectively. The number in the top-left of each result image suggests the DICE score. While KEEP effectively segments large cancerous regions, producing relatively coarse masks, it does exhibit some scattered false positives. After postprocessing with the morphological opening operation, KEEP-Post significantly reduces false positives and improves the DICE score.

We also compare KEEP with other text-based segmentation approaches, including MedSAM~\cite{ma2024medsam}, BiomedParse~\cite{zhao2025biomedparse}, and PathSeg~\cite{chen2025pathseg}, on WSI-level segmentation datasets. We simulated comparable conditions by tiling each WSI into patches at different magnifications—10×, 20×, and 40×—with a spatial size of 1120 × 1120 pixels for these segmentation models.
Among these settings, the best results were achieved at 20× magnification with 1120 × 1120 patches.
After tiling, the patches were fed into the segmentation models, and their outputs were stitched together to reconstruct full-slide masks. We then evaluated performance on three representative datasets—CAMELYON16 (48 WSIs), PANDA (a randomly selected subset of 1,000 WSIs), and AGGC (128 WSIs)—using comprehensive metrics including patch-level AUROC, AUPRC, Dice score, and ASSD.
As shown in Figure~\ref{fig:result_seg}D,E, Figure~\ref{fig:supp_seg}D-F,
and Table S3,
KEEP demonstrates significantly higher performance than other segmentation models on CAMELYON16 and AGGC, while achieving comparable AUROC and slightly lower Dice than PathSeg on PANDA. KEEP shows a higher ASSD in CAMELYON16 mainly because tumor regions occupy less than 1\% of the tissue area, making the metric highly sensitive—minor boundary shifts can cause disproportionately large distance changes.

\subsection*{KEEP enhances slide-level cancer detection }

Traditional computational methods for identifying cancerous tissues in whole slide images (WSIs) typically use multiple instance learning (MIL) for weakly supervised classification, which requires to learn the importance of informative tiles from training WSIs for cancer detection via attention mechanisms~\cite{ilse2018attention,shao2021transmil}.
The pathology foundation model CHIEF~\cite{wang2024pathology} improves upon this by combining unsupervised pre-training for extracting tile-level features with weakly supervised training for recognizing WSI patterns, achieving excellent results in various tasks. 
Despite that, these models often lack interpretability due to their need to amalgamate tens of thousands of tile-level features into a single WSI classification.

\begin{figure}
  \centering
  \includegraphics[scale=0.75]{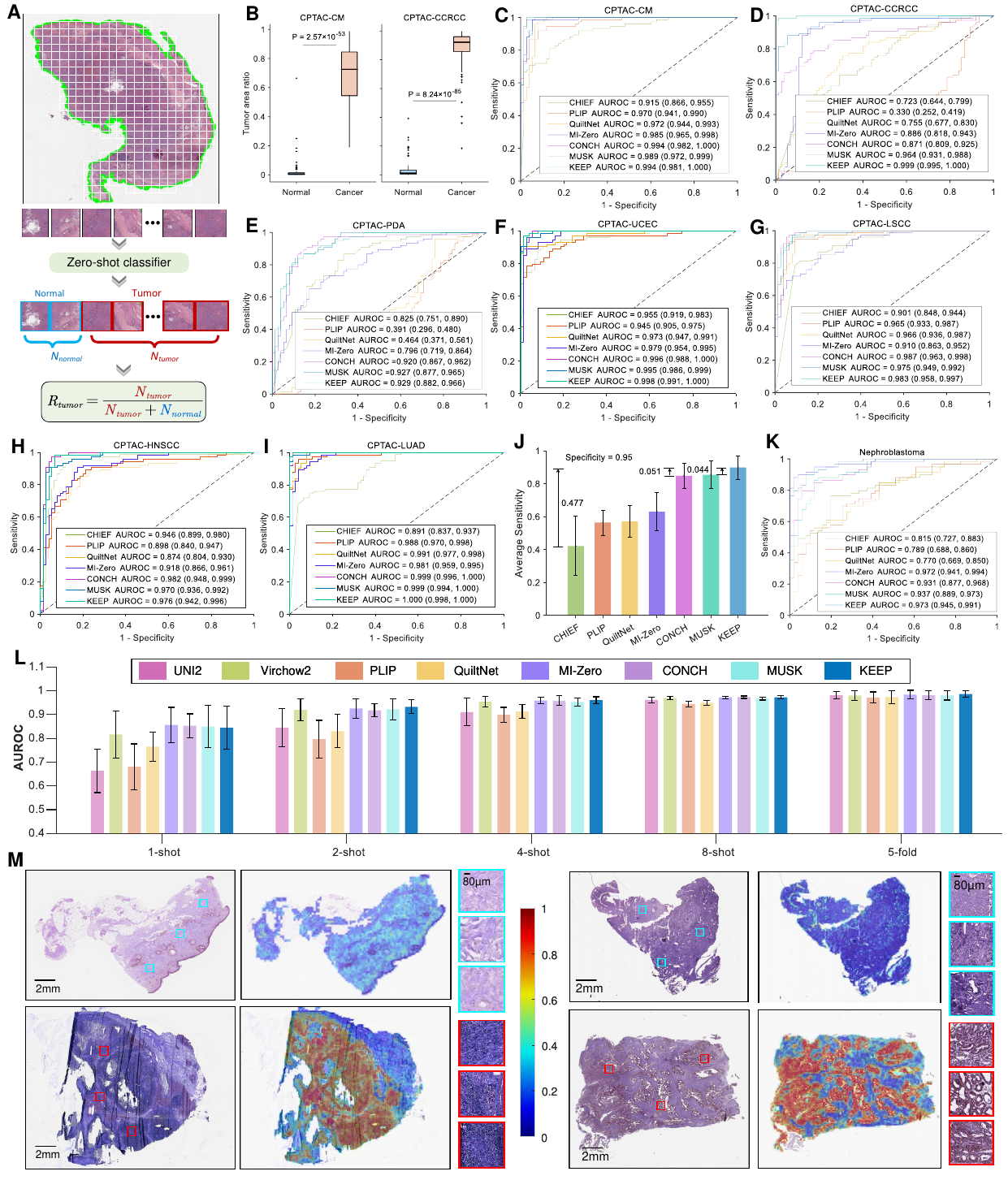}
  \caption{\textbf{KEEP enhances slide-level cancer detection}. 
  \textbf{A}. The zero-shot cancer detection scheme on WSIs, where individual tiles undergo binary classification. The probability of a slide being cancerous is determined by the predicted tumor ratio which is calculated by the ratio of tumor tiles to all valid tiles. 
  \textbf{B}. The comparison of the predicted tumor ratio between normal and cancer WSIs in CPTAC-CM and CPTAC-CCRCC datasets. Two-sided Welch’s \textit{t} test is used to assess the statistical significance of predicted tumor ratios among different WSIs. 
  \textbf{C-I}. Comparison of ROC curves across different models, including CHIEF~\cite{wang2024pathology}, PLIP~\cite{huang2023visual}, QuiltNet~\cite{ikezogwo2024quilt}, MI-Zero~\cite{lu2023visual}, CONCH~\cite{lu2024visual}, and MUSK~\cite{xiang2025vision} and KEEP, evaluated on 7 CPTAC datasets across 6 tissue anatomies: skin, kidney, pancreas, uterine, lung, and head and neck. 
  Each dataset consists of 75 normal WSIs and 75 cancer slides, 
  with each experiment using 1,000 bootstrap iterations. 
  The AUROC for each model is reported as the median along with its 95\% confidence intervals (CIs). 
  \textbf{J}. Comparison of average sensitivities across all datasets at the specificity of 0.95, the error bar denotes the standard deviation of the performance. 
  \textbf{K}. Comparison of ROC curves across different models, evaluated on a rare cancer dataset, which consists of 59 nephroblastoma WSIs and 51 normal WSIs.
  \textbf{L}. The average AUROC performance with standard deviation of different foundation models on 8 cancer detection datasets in few-shot (1,2,4,8) and 5-fold cross-validation settings.
  \textbf{M}. Example visualizations of cancer detection on CPATC-CM, CPTAC-UCEC datasets. The first and the second rows denote the normal and the cancer WSIs. The heat map is generated by the similarities between the embeddings of tile images and those of ``tumor'' prompts. Also see Figure~\ref{fig:supp_det} and Table S4.}
  \label{fig:result_det}
\end{figure}

In this paper, we explore a zero-shot setting for cancer detection, where no labeled WSIs are available for the aggregation of tile-level visual features. Practically, the zero-shot setting addresses the challenge of data scarcity that MIL-based approaches could not work effectively.
In specific, we first perform binary classification to identify cancerous tiles and aggregate these tile predictions to achieve zero-shot cancer detection. 
We compute the ratio of cancerous areas to the total tissue area (Figure~\ref{fig:result_det}A). The comparison between cancerous and normal WSIs is visualized in Figure~\ref{fig:result_det}B and Figure~\ref{fig:supp_det}A.
Our results show that the predicted tumor area ratios for cancerous WSIs differ significantly ($P < 0.001$) from that of normal slides. 
Based on this significant difference, we adopt the predicted tumor-ratio to determine whether a WSI is cancerous. Notably, in this zero-shot setting, as there is no whole slide level training, we treat each tile independently and uniformly without aggregating contextual information and different informativeness.

We evaluated on seven datasets from the Clinical Proteomic Tumor Analysis Consortium~(CPTAC\footnote{https://www.cancerimagingarchive.net/browse-collections/}), including CPTAC-CM (Cutaneous Melanoma), CPTAC-CCRCC (Clear Cell Renal Cell Carcinoma), CPTAC-PDA (Pancreatic Ductal Adenocarcinoma), CPTAC-UCEC (Uterine Corpus Endometrial Carcinoma), CPTAC-LSCC (Lung Squamous Cell Carcinoma), CPTAC-HNSCC (Head and Neck Squamous Cell Carcinoma), CPTAC-LUAD (Lung Adenocarcinoma). For each dataset, we follow the same approach as CONCH~\cite{lu2024visual}, sampling 75 cancer slide images and 75 normal slide images, and ensembling 50 text prompts for each experiment, with results presented in Figure~\ref{fig:result_det}C-I and Table S4.

We compared KEEP against established vision-language foundation models, including PLIP~\cite{huang2023conslide}, QuiltNet~\cite{ikezogwo2024quilt}, MI-Zero~\cite{lu2023visual}, CONCH~\cite{lu2024visual}, and MUSK~\cite{xiang2025vision}. Additionally, we included CHIEF~\cite{wang2024pathology} in our comparative analysis due to its demonstrated capability for zero-shot cancer detection. Vision-only foundation models such as UNI~\cite{chen2024uni} and Virchow~\cite{vorontsov2024foundation} were excluded from these evaluations as they lack the inherent zero-shot classification capabilities essential for fair comparison in our experimental framework.

Our results demonstrated that vision-language models, for example, CONCH, MUSK, and KEEP, significantly outperformed CHIEF across all datasets, underscoring that the tumor ratio metric serves as an effective way for distinguishing cancerous WSIs from normal ones (Figure~\ref{fig:result_det} and Figure~\ref{fig:supp_det}).
In particular, KEEP and CONCH achieve a median AUROC of 0.994 on skin cancer (CPTAC-CM), outperforming CHIEF by 8 points. 
For lung cancer detection (CPTAC-LUAD), KEEP, and MUSK achieve near-perfect performance (median AUROC of 1.000), while CHIEF only reaches 0.891. This substantial improvement can be attributed to the fact that, unlike the vision-only foundation model CHIEF, vision-language models such as KEEP, CONCH, and MUSK integrate predicted labels rather than embedding tile features, enabling explicit identification of cancerous regions within WSIs.

Compared to other vision-language models, KEEP achieves the best performance on five out of seven datasets. Notably, KEEP attains an average sensitivity of 0.898 at a specificity of 0.95 across all datasets, as shown in Figure~\ref{fig:result_det}J. This is more than twice the sensitivity of CHIEF, 5.1 points higher than that of CONCH, and 4.4 points higher than that of MUSK, underscoring the substantial improvement in cancer detection performance enabled by KEEP. 
In particular, on the CPTAC-CM, CPTAC-CCRCC, CPTAC-UCEC, and CPTAC-LUAD datasets, KEEP consistently exceeds a sensitivity of 0.98, as shown in Figure~\ref{fig:supp_det}B.
The similarities between predicted and ground-truth cancerous regions are visualized in heatmaps in Figure~\ref{fig:result_det}M,
further demonstrating the consistency of the model's predictions.

To further validate KEEP on rare cancer detection tasks, we collected 110 WSIs from Xinhua Hospital affiliated to Shanghai Jiao Tong University School of Medicine, including 59  nephroblastom WSIs and 51 normal WSIs. The experimental results, shown in Figure~\ref{fig:result_det}K, demonstrate that KEEP achieves comparable AUROC with MI-Zero, which outperforms CHIEF by about 15 points, indicating the advantage of vision-language models in zero-shot rare cancer detection tasks.

To better assess the practical applicability in real clinical scenarios, we conducted additional few-shot (1,2,4,8-shot) and 5-fold cross-validation experiments on all cancer detection benchmarks. 
Specifically, we compared the advanced vision-only foundation models (Virchow2~\cite{zimmermann2024virchow2} and UNI2~\cite{chen2024uni}) and vision-language foundation models (PLIP, QuiltNet, MI-Zero, CONCH, MUSK, and KEEP) using TransMIL~\cite{shao2021transmil} aggregation strategies under few-shot and 5-fold settings. 
The results are shown in Figure~\ref{fig:result_det}L and Table S4.
It can be seen that KEEP’s zero-shot accuracy not only surpasses that of all other vision-language models under the zero-shot setting, but also exceeds the few-shot results (1–8shot) of all foundation models, demonstrating the strong generalization capability of our model without any task-specific tuning.  
Furthermore, when sufficient fine-tuning data become available, model performance can indeed be further improved. Under the 5-fold setting, KEEP again achieves the best overall performance (AUROC=0.987$\pm$0.015) in cancer detection tasks, confirming that KEEP benefits from additional supervision while maintaining excellent baseline generalization.

\subsection*{KEEP enhances slide-level cancer subtyping}

\begin{figure}
  \centering
  \includegraphics[scale=0.72]{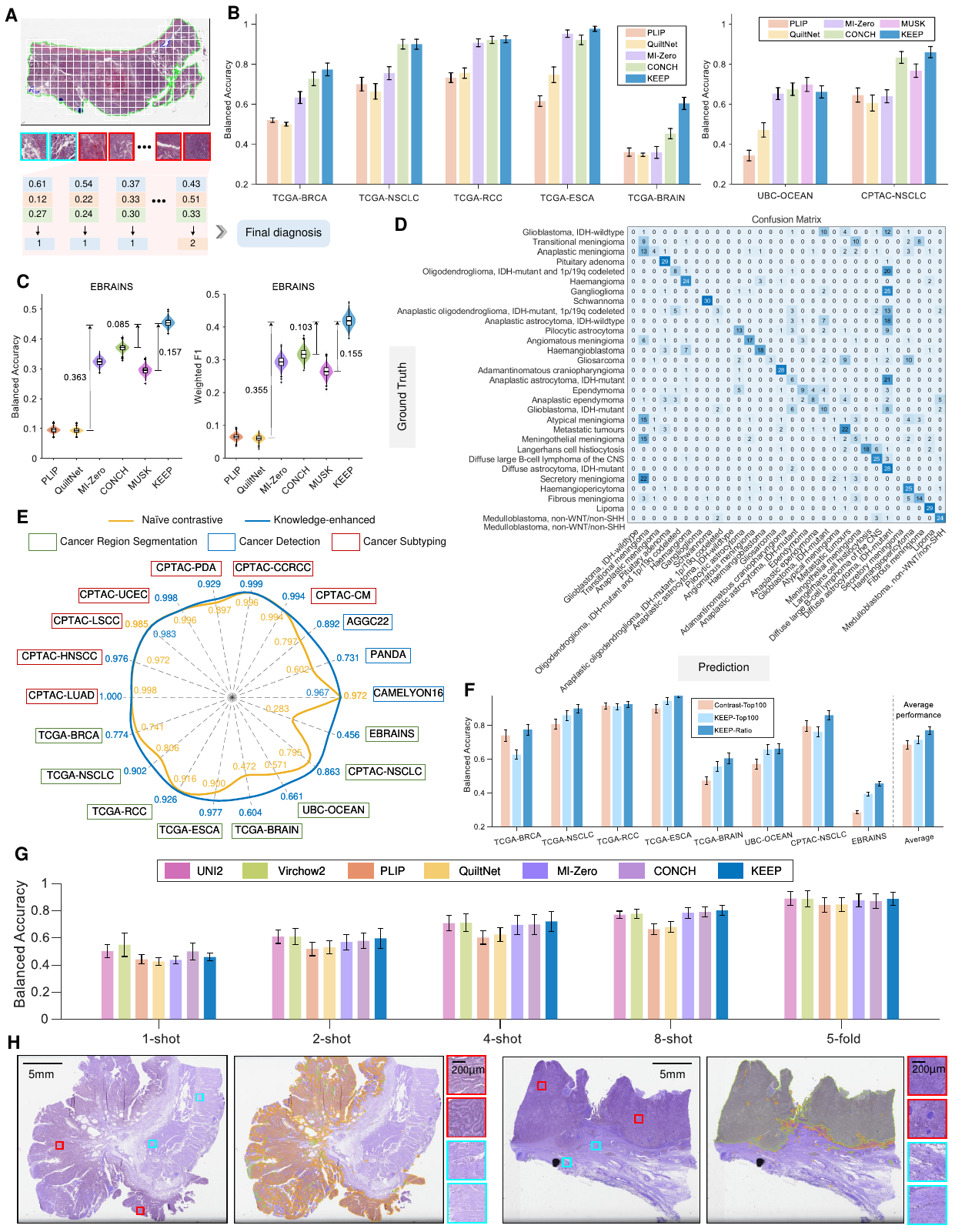}
  \vspace{-10pt}
  \caption{\textbf{KEEP enhances slide-level cancer subtyping}. 
  \textbf{A}. The zero-shot cancer subtyping scheme on WSIs, where individual tiles undergo multi-class classification, including a ``normal'' label and tumor subtype labels. The probability of a slide being classified as type I is determined by the ratio of type I tiles to all valid tiles. 
  \textbf{B}. Comparison of average balanced accuracy with standard deviation across different models on seven datasets with common cancer subtypes, with each experiment using 1,000 bootstrap iterations. 
  \textbf{C}. Performance comparison of different models on the rare cancer subtyping dataset, EBRAINS, which consists of 30 rare brain cancer subtypes, each with 30 WSIs. \textbf{D}. The confusion matrix of the KEEP model on the rare brain cancer dataset, EBRAINS. 
  \textbf{E}. Ablation results on WSI tasks. Performance comparison between na\"ive contrastive and knowledge-enhanced (KEEP).
  \textbf{F}. Ablation results between na\"ive contrastive with Top-100 pooling strategy (Contrast-Top100), KEEP with Top-100 pooling strategy (KEEP-top100) and KEEP with tumor-ratio strategy (KEEP-Ratio).
  \textbf{G}. The average BACC performance with standard deviation of different foundation models on 7 common cancer subtyping datasets in few-shot (1,2,4,8) and 5-fold cross-validation settings.
\textbf{H}. Example WSIs for tumor subtyping. The left and the right WSIs denote esophagus adenocarcinoma and esophagus squamous cell carcinoma, respectively. The orange and the green masks denote the predicted regions of adenocarcinoma and squamous cell carcinoma, respectively. The blue squares denote the tile image from the area with normal predictions. (H)Also see Figure~\ref{fig:supp_sub} and Table S5.}
  \label{fig:result_sub}
\end{figure}

Identifying tumor subtypes is essential for accurate cancer diagnosis and personalized treatment. Existing models, including multiple instance learning (MIL)-based approaches and the foundation model Prov-GigaPath~\cite{xu2024whole}, aggregate image features from individual tiles into a WSI-level representation for multi-class classification. 
However, these methods require a large number of labeled whole slide images for training, which limits their scalability to new cancer types.

Pathology vision-language models offer a promising zero-shot paradigm, 
where the predicted labels of individual tile images are aggregated, 
rather than their features, to determine the final classification. For example, MI-Zero~\cite{lu2023visual} and CONCH~\cite{lu2024visual} predict the subtype probability for each tile and then integrate the top-K predictions to classify the entire WSI. 
This zero-shot approach is highly adaptable and can be easily extended to new datasets.

To enhance this paradigm in a non-parametric manner, we adopt the same approach as cancer detection that uses the ratio of cancer subtype areas to total tissue areas as the subtype probability, as shown in Figure~\ref{fig:result_sub}A. Compared to the top-K strategy, the ratio-based approach provides a more rigorous estimate for the prediction of subtyping probabilities. The mathematical details can be found in Methods.
We evaluate the performance of different models on both common and rare cancer subtypes. 
The common cancer subtyping tasks include 7 whole slide image (WSI) datasets from The Cancer Genome Atlas (TCGA\footnote{https://portal.gdc.cancer.gov/}), CPTAC, and other resources: 
TCGA-BRCA (invasive breast carcinoma), TCGA-NSCLC (non-small-cell lung carcinoma), TCGA-RCC (renal cell carcinoma), TCGA-ESCA (esophagus carcinoma), TCGA-Brain (brain cancer), UBC-OCEAN~\cite{farahani2022ubc,asadi2024ubc2} (ovarian cancer), and CPTAC-NSCLC (non-small-cell lung carcinoma). 
Specifically, TCGA-BRCA consists of two subtypes: 
invasive ductal carcinoma (IDC) and invasive lobular carcinoma (ILC). 
TCGA-NSCLC contains lung adenocarcinoma (LUAD) and lung squamous cell carcinoma (LUSC). 
TCGA-RCC is divided into chromophobe renal cell carcinoma (CHRCC), clear-cell renal cell carcinoma (CCRCC), and papillary renal cell carcinoma (PRCC).
TCGA-ESCA includes two subtypes: squamous cell carcinoma and adenocarcinoma. 
TCGA-BRAIN consists of three subtypes: glioblastoma, astrocytoma, and oligodendroglioma. 
UBC-OCEAN contains five subtypes: ovarian clear cell carcinoma (CC), ovary endometrioid carcinoma (EC), high-grade ovary serous carcinoma (HGSC), low-grade ovary serous carcinoma (LGSC), ovarian mucinous carcinoma (MC). 
For the datasets collected from TCGA and CPTAC, we follow CONCH~\cite{lu2024visual} to randomly sample 75 WSIs for each cancer subtype (65 for ESCA due to the number of original whole cases). For UBC-OCEAN, we randomly sample 35 WSIs for each subtype. The details of all the above datasets are listed in Table S1.

Following CONCH, we ensemble 50 text prompts for each task and present the results in Figure~\ref{fig:result_sub}B (MUSK is excluded from the evaluation on TCGA datasets as it was trained on these datasets.), 
Figure~\ref{fig:supp_sub}A,
and Table S5.
Notably, KEEP outperforms other models on 6/7 datasets. 
For the brain cancer subtyping task, KEEP achieves an average balanced accuracy of 0.604, which is 0.15 higher than CONCH and 0.25 higher than the other models.

For rare cancer subtyping, we evaluate different models on the EBRAINS~\cite{roetzer2022ebrains} dataset, which includes 128 subtypes of rare brain tumors. In accordance with CONCH, we select 30 rare subtypes, each containing more than 30 whole slide images (WSIs), for cancer subtyping evaluation. The results (Figure~\ref{fig:result_sub}C) show that KEEP achieves a median balanced accuracy of 0.456, more than four times higher than PLIP and QuiltNet, 8.5 points higher than CONCH, and 15.5 points higher than MUSK. Figure~\ref{fig:result_sub}D displays the confusion matrix for KEEP on the EBRAINS dataset, where it performs particularly well on subtypes such as schwannoma, adamantinomatous craniopharyngioma, and lipoma. 
However, certain subtypes exhibit notable misclassification patterns. In particular, we observed confusion between `GBM, IDH-wildtype' and `GBM, IDH-mutant' pairs, as well as between `anaplastic astrocytoma, IDH-wildtype' and `anaplastic astrocytoma, IDH-mutant' variants. This is primarily because the morphological phenotypes associated with IDH mutation status remain poorly defined and inadequately documented, consequently limiting the availability of high-quality training data necessary for vision-language models to perform reliable molecular subtyping. To enhance the molecular subtyping performance on EBRAINS, we implemented TransMIL~\cite{shao2021transmil} for aggregating tile-level features across 5-shot and 10-shot fine-tuning scenarios. The experimental results, presented in Table S6, demonstrate that KEEP's visual representations outperformed all competing models, exhibiting the BACC of 0.648 across 30 subtypes under the 10-shot configuration (19.2-point enhancement compared to its zero-shot performance).

To validate the enhancement of disease knowledge, we compare the performance of KEEP (knowledge-enhanced) and that of na\"ive contrastive baseline (same backbone, without knowledge enhancement) on all WSI tasks, 
including zero-shot cancer region segmentation, cancer detection, and cancer subtyping. 
The experimental results (Figure~\ref{fig:result_sub}E, Figure~\ref{fig:supp_sub},
and Table S5) show that the knowledge-enhanced model achieves significant improvement in 16/18 WSI datasets over the na\"ive contrastive baseline. In particular, for cancer region segmentation tasks, knowledge-enhanced models achieved nearly 10\% and 12.9\% higher AUROC than the na\"ive contrastive strategy on PANDA and AGGC22 datasets. 
For the cancer detection datasets, 
KEEP demonstrates better performance over the na\"ive contrastive approach on 6/7 cancer detection benchmarks. 
For the cancer subtyping datasets,
KEEP outperforms the non-knowledge baseline on all cancer subtyping benchmarks.

Figure~\ref{fig:result_sub}F and Table S5
exhibit the detailed ablation results of knowledge enhancement (Contrastive-Top100 \textit{vs} KEEP-Top100) as well as the ratio-based strategy (KEEP-Top100 \textit{vs} KEEP-Ratio), where Top100 represents top-100 pooling strategy developed by MI-Zero and CONCH.
It can be seen that KEEP-Top100 achieves comparable or better performance than Contrastive-Top100 in 6 out of 8 datasets, with a significant improvement of 11 points on the rare tumor dataset, EBRAINS, indicating that the disease knowledge can substantially facilitate the zero-shot performance of rare tumor subtyping tasks. 
The performance comparison between KEEP-Top100 and KEEP-Ratio, presented in Figure~\ref{fig:result_sub}F and Figure~\ref{fig:supp_sub}E,
show that our subtype ratio strategy outperforms Top100 pooling on all datasets. Specifically, KEEP-ratio achieves an average balanced accuracy of 0.860 on CPTAC-NSCLC and 0.774 on TCGA-BRCA, surpassing KEEP-top100 by 0.10 and 0.15, respectively. 
All the ablation details and additional results on semantic groups, caption augmentation, loss function, knowledge graph, text encoder, and data scale can be found in Supplementary Note 1, Figure~\ref{fig:supp_sub}, and Figure Table S5.

To evaluate the performance of our model after fine-tuning, we conducted additional few-shot (1, 2, 4, and 8-shot) and five-fold cross-validation experiments across seven representative cancer subtyping benchmarks.
We compared state-of-the-art vision-only foundation models (Virchow2 and UNI2) with vision–language foundation models (PLIP, QuiltNet, MI-Zero, CONCH, MUSK, and KEEP) using the TransMIL aggregation framework.
KEEP exhibited the best performance among vision–language models under both zero-shot and few-shot settings, even surpassing the few-shot (1–8 shot) results of all other foundation models (Table S5), indicating robust generalization without task-specific adaptation.
With increasing amounts of fine-tuning data, model accuracy further improved. Under the five-fold setting, KEEP consistently outperformed other vision–language models (Figure~\ref{fig:result_sub}G) and achieved performance comparable to UNI2 (0.890 and 0.888), demonstrating that additional supervision can enhance accuracy while maintaining strong baseline generalization.

Figure~\ref{fig:result_sub}H
shows the visualization of the semantic segmentation of cancer subtypes, combining predicted labels with identified cancerous regions. These visualizations underscore the exceptional interpretability of our approach in cancer subtyping tasks.

\subsection*{KEEP enhances rare cancer subtyping}

\begin{figure}[!t]
  \centering
  \includegraphics[scale=0.8]{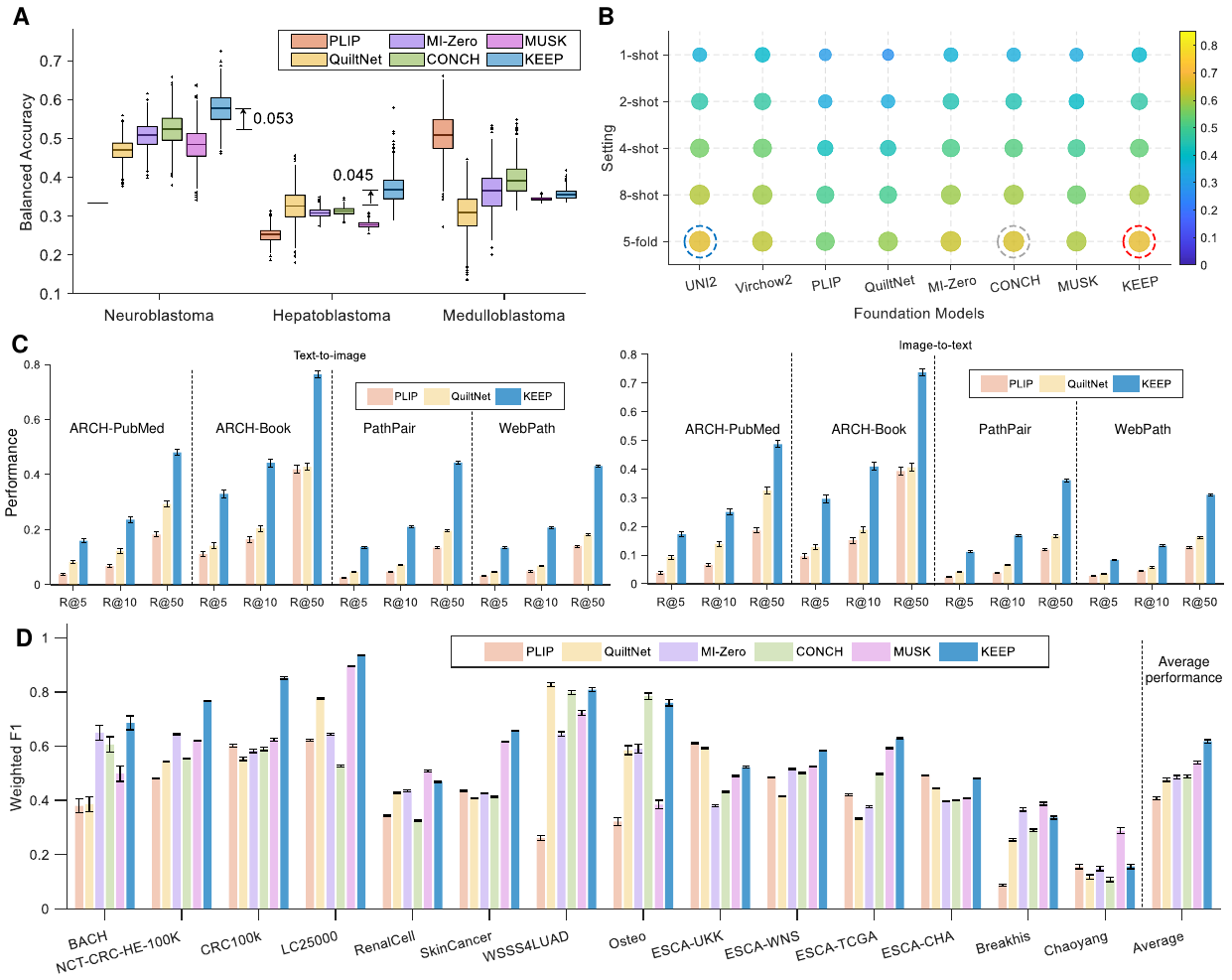}
  \caption{\textbf{KEEP enhances rare pediatric cancer subtyping and zero-shot pathology tile image profiling}. 
  \textbf{A}. Zero-shot performance comparison of different models on the clinical rare cancer subtyping task.
  \textbf{B}. The average BACC performance of different foundation models on 4 rare cancer subtyping datasets, including EBRAINS, Neuroblastoma, Hepatoblastoma, and Medulloblastoma, in few-shot (1,2,4,8) and 5-fold cross-validation settings.
  \textbf{C}. Performance comparison of different models on the cross-modal retrieval task. R@K (k = 5, 10, and 50) denotes Recall@K, the ratio of correctly retrieved queries in Top-K retrieved samples. 
  \textbf{D}. Performance comparisons of different models on the zero-shot tile image classification task. The error bar denotes the standard deviation of the results from 1000 bootstrapping iterations. `Average' suggests the mean of performance across all datasets. Also see Figure~\ref{fig:supp_xinhua} and Table S6.}
  \label{fig:result_tile}
\end{figure}

To further validate the KEEP model on a broader spectrum of rare cancers, we also collected 816 WSIs of pediatric cancer patients from Xinhua Hospital affiliated to Shanghai Jiao Tong University School of Medicine. Specifically, the collection includes: 
\textbf{(i) Neuroblastoma}: 136 WSIs covering three clinically distinct subtypes, including 33 for ``Ganglioneuroblastoma, intermixed'',
33 for ``Differentiating neuroblastoma'', and 70 for ``Poorly differentiated neuroblastoma'';
\textbf{(ii) Hepatoblastoma}: 442 WSIs covering four subtypes, including 10 for ``Epithelial macrotrabecular pattern of hepatoblastoma'', 76 for ``Mixed epithelial and mesenchymal hepatoblastoma'', 177 for ``Epithelial mixed fetal and embryonal hepatoblastoma'', and 179 for ``Pure fetal hepatoblastoma with low mitotic activity''.
\textbf{(iii) Medulloblastoma}: 238 WSIs covering three subtypes, including 11 ``Large Cell/Anaplastic medulloblastoma'', 30 for ``Desmoplastic nodular medulloblastoma'', and 197 for ``Classic medulloblastoma''.

We conducted both zero-shot and fine-tuning experiments on KEEP, PLIP~\cite{huang2023visual}, QuiltNet~\cite{ikezogwo2024quilt}, MI-Zero~\cite{lu2023visual}, CONCH~\cite{lu2024visual}, and MUSK~\cite{xiang2025vision}. The results are summarized in Figure~\ref{fig:result_tile}A,B. 
For zero-shot evaluation, the performance in Figure~\ref{fig:result_tile}A and Table S6
demonstrates that KEEP significantly outperforms competing models in the subtyping tasks of neuroblastoma and hepatoblastoma. Although KEEP appears to perform less effectively on the medulloblastoma dataset, Figure~\ref{fig:supp_xinhua}
reveals that all models actually show very limited discriminative ability when identifying medulloblastoma subtypes.
In fact, each model predominantly categorizes the majority of samples into one subtype, indicating the inherent challenges this particular classification task presents.

For fine-tuning experiments, we adopt the visual encoder of each model as the visual backbone, fine-tune it with few-shot (1,2,4,8-shot) and five-fold cross-validation settings using TrainsMIL~\cite{shao2021transmil}. To mitigate potential data bias and ensure robust performance assessment, we conducted each few-shot experiment across 10 independent iterations, with each iteration employing a different randomized training dataset.
The experimental results on 4 rare cancer datasets, including EBRAINS, Neuroblastoma, Hepatoblastoma, and Medulloblastoma, shown in Figure~\ref{fig:result_tile}B and Table S6,
suggest that KEEP consistently outperforms all other vision–language models across all few‑shot scenarios, while achieving the second‑best performance in the 1‑shot, 2‑shot, and 8‑shot settings. This demonstrates its robust adaptability even when minimal fine‑tuning data are available. Furthermore, as more labeled data become available, all models benefit from additional supervision, yet KEEP maintains its leading advantage, achieving the best overall performance under the 5‑fold setting (BACC=0.671$\pm$0.085).

\subsection*{KEEP enhances zero-shot pathology tile image profiling}

In this section, we conduct an evaluation of KEEP on tile-level tasks, including cross-modal retrieval and zero-shot tile image classification. 

\noindent\textbf{Cross-modal Retrieval.} 
We evaluate on 4 pathology image-text datasets: ARCH-PubMed~\cite{gamper2021arch}, ARCH-Book~\cite{gamper2021arch}, Pathpair~\cite{zhou2024kep}, and WebPath, a dataset collected from professional websites. As MI-Zero and CONCH do not release their image-text training data, and the training dataset of MUSK involves PubMed papers and textbooks,
to avoid unfair comparison from data leakage, we limit our comparison to PLIP and QuiltNet. The experimental results (Figure~\ref{fig:result_tile}C), 
show that KEEP outperforms PLIP and QuiltNet by a significant margin across all datasets, in both text-to-image and image-to-text retrieval tasks.
This improvement is attributed to the well-trained visual encoder and the knowledge-enhanced text encoder. The comparison with MI-Zero, CONCH, and MUSK is shown in Table S6.

\noindent\textbf{Tile Image Classification.}
For the tile image classification task, 
we compile 14 datasets covering seven human tissue types: breast~(BACH~\cite{aresta2019bach} and Breakhis~\cite{spanhol2015breakhis}), 
colon~(NCT-CRC-HE-100K~\cite{kather2018nct100k}, CRC100K~\cite{kather2019crc100k}, and Chaoyang~\cite{zhu2021chaoyang}), lung~(LC25000~\cite{borkowski2019lc25000} and WSSS4LUAD~\cite{han2022wsss4luad}), 
kidney~(RenalCell~\cite{brummer2022renalcell}), 
bone~(Osteo~\cite{arunachalam2019osteo}), 
skin~(SkinCancer~\cite{kriegsmann2022skincancer}), 
and esophagus (ESCA-UKK, ESCA-WNS, ESCA-TCGA, and ESCA-CHA~\cite{tolkach2023esca}). The number of classes per dataset ranges from 2 to 16. Full dataset details are provided in the Methods section and Table S1.

In accordance with PLIP, we concatenate a template phrase with the class names (\textit{a histopathology image of} \{class name\}) in each dataset to construct the text prompts for zero-shot tile classification. The bootstrapping performance of different models is shown in Figure~\ref{fig:result_tile}D and Table S6
Notably, KEEP achieves the best performance on 7 out of 14 datasets. In particular, KEEP achieves the best average performance across all datasets, which is 0.078 and 0.130 higher than that of MUSK and CONCH, respectively.

\section*{DISCUSSION}

In this study, we present KEEP, a novel vision-language foundation model specifically designed to tackle challenges in computational pathology. By incorporating disease-specific knowledge, KEEP achieves state-of-the-art performance in zero-shot cancer diagnosis. 
In particular, for \textbf{cancer detection}, KEEP significantly outperforms CHIEF, achieving a notable improvement in sensitivity (0.898) at a specificity of 0.95 across multiple cancer types. Similarly, in \textbf{cancer subtyping}, KEEP outperforms CONCH and MUSK by integrating disease-specific knowledge, which improves the alignment between pathology images and subtype semantics. Notably, in \textbf{rare cancer subtyping tasks}, KEEP achieves a balanced accuracy improvement of 8.5 points over CONCH on EBRAINS. Furthermore, when applied to in-house rare pediatric cancer diagnostic tasks, KEEP outperforms other foundation models in the detection task of nephroblastom dataset, and the subtyping tasks on the neuroblastoma and hepatoblastoma datasets.

The superior performance of KEEP compared to other models can be primarily attributed to two factors: the injection of disease knowledge during training and the tumor-ratio-based prediction strategy employed in downstream tasks:
\textbf{(i)} integration of disease knowledge during training: disease knowledge is incorporated through three key mechanisms. First, the language model is employed to encode the disease knowledge graph (KG), aligning the representation space of disease names, definitions, synonyms, and hierarchical relationships. This alignment serves as a bridge, implicitly linking pathology images with their corresponding disease types during the vision-language pre-training process. Second, the extensive use of disease synonyms within the disease KG explicitly highlights critical disease-related information in the textual descriptions of images. This transformation converts weak supervision signals from free-text annotations into stronger disease-level supervision, thereby reinforcing the connection between pathology images and disease entities. Finally, the hierarchical structure of the disease KG organizes image-text pairs into semantic groups with hypernym-hyponym relationships, significantly enhancing alignment accuracy.
\textbf{(ii)} tumor-ratio-based prediction in downstream tasks: 
for both cancer detection and subtyping, mathematical analysis and experimental results demonstrate that the tumor-ratio-based prediction serves as an intuitive and effective way for estimating the neoplastic and subtyping probability of WSIs.
Unlike methods that aggregate tile-level features, the tumor-ratio-based approach leverages tumor region localization to deliver superior interpretability, a critical requirement for clinical applications. Moreover, compared to approaches that predict tumor subtypes by selecting top-k tumor patches, the tumor-ratio-based method is non-parametric, offering a more straightforward and transparent classification process. 

In conclusion, our results demonstrate that KEEP offers a promising research tool for computational pathology by injecting domain-specific knowledge into vision-language models. KEEP shows potential for advancing cancer diagnosis research, particularly in zero-shot scenarios and rare cancer diagnosis, where traditional approaches face data limitations. However, clinical deployment will require further development including larger-scale validation studies, task-specific optimization, and integration with existing clinical workflows to meet the reliability standards necessary for real-world diagnostic applications.

\subsection*{Limitations of the study}


While promising, this study also faces several limitations:
\textbf{(i)} although KEEP demonstrates strong performance in rare cancer subtyping, its predictions for certain subtypes remain limited due to the scarcity of these cases in the image-text training data. 
In these instances, few-shot learning, where at least one whole slide image (WSI) per subtype is available, could enhance model performance by capturing the diversity within rare cancer subtypes.
\textbf{(ii)} while pathology vision-language models exhibit robust zero-shot ability, their performance is often dependent on prompt engineering, which can limit their adaptability and robustness across diverse datasets. 
A promising avenue for improvement is prompt learning, wherein the model learns a trainable prompt from a small set of example WSIs, replacing manually designed prompts. This approach would enable the model to adapt more effectively to varied datasets and tasks, enhancing its generalizability and robustness. 
\textbf{(iii)} another promising avenue for improvement is the multi-modal alignment that introduces genomic~\cite{xu2024multimodal}, \cite{vaidya2025molecular} or epigenomic~\cite{hoang2024prediction} information.

\newpage







\section*{RESOURCE AVAILABILITY}


\subsection*{Lead contact}


Requests for further information and resources should be directed to and will be fulfilled by the lead contact, Weidi Xie (weidi@sjtu.edu.cn).

\subsection*{Materials availability}


This study did not generate new materials.

\subsection*{Data and code availability}



\begin{itemize}
    \item The disease Knowledge, including disease entities and hypernym relations are available in Disease Ontology (DO)~(\url{https://disease-ontology.org/do/}). 
    Disease synonyms and definitions are available in Unified Medical Language System (UMLS)~(\url{http://www.nlm.nih.gov/research/umls/licensedcontent/umlsknowledgesources.html}). 
    The pathology image-text pairs for alignment are available in OpenPath~(\url{https://huggingface.co/vinid/plip}) and Quilt1M~(\url{https://github.com/wisdomikezogwo/quilt1m}). Test datasets for cancer region segmentation are available in CAMELYON16~(\url{https://camelyon16.grand-challenge.org/}), PANDA~(\url{https://panda.grand-challenge.org/data/}), and AGGC22~(\url{https://aggc22.grand-challenge.org/}). For cancer detection, test datasets of CPTAC-CM, CPTAC-CCRCC, CPTAC-PDA, CPTAC-UCEC, CPTAC-LSCC, CPTAC-HNSCC, CPTAC-LUAD are available in CPTAC~(\url{https://proteomics.cancer.gov/programs/cptac}). For cancer subtyping, test datasets of TCGA-BRCA, TCGA-NSCLC, TCGA-RCC, TCGA-ESCA, TCGA-BRAIN are available in TCGA~(\url{https://portal.gdc.cancer.gov/}). Other datasets for cancer subtyping are available in CPTAC-NSCLC~(\url{https://proteomics.cancer.gov/programs/cptac}), EBRAINS~(\url{https://data-proxy.ebrains.eu/datasets/}), and UBC-OCEAN ~(\url{https://www.kaggle.com/competitions/UBC-OCEAN/}). The sources of all tile datasets are listed in Table S1. The WSIs obtained from Xinhua Hospital for model testing and the associated de-identified clinical information are available at ~\href{https://huggingface.co/datasets/Firehdx233/KidRare}{KidRare} upon reasonable request and are subject to an application/approval process for scientific research purposes only.
    \item The source codes for KEEP are available at~\href{https://github.com/MAGIC-AI4Med/KEEP}{MAGIC-AI4Med/KEEP} under the MIT License.
    \item The pre-trained KEEP weights and model card are accessible on Hugging Face at~\href{https://huggingface.co/Astaxanthin/KEEP}{Astaxanthin/KEEP} under the MIT License.
    \item The curated pathology image–text semantic groups, disease knowledge graph, and annotated data used for KEEP are also available at~\href{https://huggingface.co/datasets/Loie/KEEP_dataset}{KEEP Dataset} under the MIT License.
\end{itemize}

\section*{ACKNOWLEDGMENTS}


{\small This work is supported by the National Key R\&D Program of China (No. 2022ZD0161400), the Scientific Research Innovation Capability Support Project for Young Faculty (ZYGXQNJSKYCXNLZCXM-I22), the National Natural Science Foundation of China (No. 24Z031503678), and China Postdoctoral Science Foundation (Certificate Number: 2023M741850). 
}

\section*{AUTHOR CONTRIBUTIONS}


X.Z., L.S., W.X. developed the research concept and experimental design. X.Z., L.S., and D.H. conducted all experiments and provided result analysis. 
For in-house rare cancer data collection and experiments: W.G., R.W. and L.W. conducted comprehensive examination and scanning the pathology whole slides from Xinhua hospital. X.Y and X.S. collected and verified the associated patient clinical information, ensuring data accuracy and completeness. They were also responsible for obtaining institutional review board (IRB) approval and ensuring compliance with ethical guidelines for patient data usage. G.W. and W.G. provided independent manual annotations for tumor regions on WSIs. W.X., Y.W., and K.S. provided comprehensive leadership and management for the entire project. Y.Z., W.X., and Y.W. directed the sophisticated modeling components and algorithmic development, while K.S. expertly coordinated all hospital-related aspects, including systematic data collection, rigorous manual annotation processes, and valuable clinical insights that enhanced the project's medical relevance and applicability. The manuscript was drafted by X.Z., L.S. and W.X. with critical revisions from all co-authors.


\section*{DECLARATION OF INTERESTS}


The authors declare no competing interests.




\section*{SUPPLEMENTAL INFORMATION INDEX}



{\small
\begin{description}
  \item Document S1. Supplementary Note 1, related to Figure 5, Supplementary Note 2, related to STAR Methods.
  \item Table S1. Description and access information for datasets used in this study. Related to Figure 1.
  \item Table S3. Detailed experimental results for cancer region segmentation, related to Figure 3.
  \item Table S4. Detailed experimental results for cancer detection, related to Figure 4.
  \item Table S5. Detailed experimental results for cancer subtyping and ablation study, related to Figure 5.
  \item Table S6. Detailed experimental results for rare cancer subtyping and zero-shot pathology tile image profiling, related to Figure 6.
\end{description}
}
\newpage








\newpage

\newpage


\bibliography{references}

\bigskip


\newpage



\section*{STAR METHODS}







\subsection*{Dataset description}
\noindent\textbf{Cancer Cell Segmentation.}
We collected three WSI datasets, including Camelyon16~\cite{bejnordi2017camelyon16}, PANDA~\cite{bulten2022panda}, and AGGC22~\cite{huo2024aggc22}. The test dataset in Camelyon16 contains 48 WSIs from breast cancer tissue with cancer cell masks. PANDA consists of 10,494 WSIs from prostate cancer tissue. The test dataset in AGGC22 contains 128 WSIs from prostate tumor tissue with annotated masks. 

\noindent\textbf{Cancer Detection.}
We conduct cancer detection on seven WSI datasets, including CPTAC-CM (Cutaneous Melanoma), CPTAC-CCRCC (Clear Cell Renal Cell Carcinoma), CPTAC-PDA (Pancreatic Ductal Adenocarcinoma), CPTAC-UCEC (Uterine Corpus Endometrial Carcinoma), CPTAC-LSCC (Lung Squamous Cell Carcinoma), CPTAC-HNSCC (Head and Neck Squamous Cell Carcinoma), CPTAC-LUAD (Lung Adenocarcinoma). For each dataset, we follow CONCH~\cite{lu2024visual} to randomly sample 75 cancer slide images and 75 normal slide images. 
To further validate KEEP on rare cancer detection tasks, we collected 110 WSIs from Xinhua Hospital affiliated to Shanghai Jiao Tong University School of Medicine, including 59 \textbf{nephroblastom} WSIs and 51 normal WSIs. All the rare cancer data collected from the hospital were approved by Ethics Committee of Xinhua Hospital Affiliated to Shanghai Jiaotong University School of Medicine (XHEC-C-2025-015-1).

\noindent\textbf{Cancer Subtyping.}
We conduct cancer detection on eight WSI datasets, including TCGA-BRCA, TCGA-LUNG (including TCGA-LUAD and TCGA-LUSC), TCGA-RCC (including TCGA-KICH, TCGA-KIRP, TCGA-KIRC), TCGA-ESCA, TCGA-BRAIN (TCGA-LGG and TCGA-GBM), UBC-OCEAN, CPTAC-LUNG (CPTAC-LUAD and CPTAC-LSCC), and EBRAINS. 
Specifically, TCGA-BRCA consists of two subtypes: invasive ductal carcinoma (IDC) and invasive lobular carcinoma (ILC). TCGA-NSCLC contains lung adenocarcinoma (LUAD) and lung squamous cell carcinoma (LUSC). TCGA-RCC is divided into chromophobe renal cell carcinoma (CHRCC), clear-cell renal cell carcinoma (CCRCC), and papillary renal cell carcinoma (PRCC).
TCGA-ESCA includes two subtypes: squamous cell carcinoma and adenocarcinoma, NOS. TCGA-BRAIN consists of three subtypes: glioblastoma, astrocytoma, NOS, and oligodendroglioma, NOS. UBC-OCEAN contains five subtypes: ovarian clear cell carcinoma (CC), ovary endometrioid carcinoma (EC), high-grade ovary serous carcinoma (HGSC), low-grade ovary serous carcinoma (LGSC), ovarian mucinous carcinoma (MC). EBRAINS is composed of 128 subtypes of common and rare brain tumors. The details of all the above datasets are listed in Table S1.
For the datasets collected from TCGA and CPTAC, we follow CONCH~\cite{lu2024visual} to randomly sample 75 WSIs for each cancer subtype (65 for ESCA due to the number of original whole cases). While for EBRAINS, we follow CONCH to use the subtypes with more than 30 WSIs and finally obtain 900 WSIs in total. 
We also collected 816 WSIs of pediatric cancer patients from Xinhua Hospital affiliated to Shanghai Jiao Tong University School of Medicine. Specifically, the collection includes: 
\textbf{(i) Neuroblastoma}: 136 WSIs covering three clinically distinct subtypes, including 33 for ``Ganglioneuroblastoma, intermixed'',
33 for ``Differentiating neuroblastoma'', and 70 for ``Poorly differentiated neuroblastoma'';
\textbf{(ii) Hepatoblastoma}: 442 WSIs covering four subtypes, including 10 for ``Epithelial macrotrabecular pattern of hepatoblastoma'', 76 for ``Mixed epithelial and mesenchymal hepatoblastoma'', 177 for ``Epithelial mixed fetal and embryonal hepatoblastoma'', and 179 for ``Pure fetal hepatoblastoma with low mitotic activity''.
\textbf{(iii) Medulloblastoma}: 238 WSIs covering three subtypes, including 11 ``Large Cell/Anaplastic medulloblastoma'', 30 for ``Desmoplastic nodular medulloblastoma'', and 197 for ``Classic medulloblastoma''.
All the rare cancer data collected from the hospital were approved by Ethics Committee of Xinhua Hospital Affiliated to Shanghai Jiaotong University School of Medicine (XHEC-C-2025-015-1).

\noindent\textbf{Tile-level Image Classification.}
We collect 14 tile-level pathology image datasets covering 8 human tissues for the classification task, including BACH~\cite{aresta2019bach}, NCT-CRC-HE-100K~\cite{kather2018nct100k}, CRC100K~\cite{kather2019crc100k}, LC25000~\cite{borkowski2019lc25000}, RenalCell~\cite{brummer2022renalcell}, SkinCancer~\cite{kriegsmann2022skincancer}, WSSS4LUAD~\cite{han2022wsss4luad}, Osteo~\cite{arunachalam2019osteo}, ESCA\_UKK, ESCA\_WNS, ESCA\_TCGA, ESCA\_CHA~\cite{tolkach2023esca}, 
Breakhis~\cite{spanhol2015breakhis}, Chaoyang~\cite{zhu2021chaoyang}. Each dataset contains multiple types (ranging from 2 to 16) of H\&E stained cell micrographs. The details of each dataset are shown in Table S1.

\noindent\textbf{Cross-modal Retrieval.} 
Cross-modal retrieval aims to retrieve the correct caption for a given image and vice versa. We collect three pathology image-text datasets for the cross-modal retrieval task, including ARCH~\cite{gamper2021arch}, PathPair~\cite{zhou2024kep}, and WebPath collected from public websites. For the ARCH dataset, we follow PLIP~\cite{huang2023visual} to split the ARCH~\cite{gamper2021arch} dataset into ARCH-PubMed and ARCH-book.

\subsection*{Method details}


\subsubsection*{Disease Knowledge Graph Construction}

We construct a comprehensive disease knowledge graph by integrating data from publicly available databases: the Disease Ontology (DO)~\cite{schriml2012disease} and the Unified Medical Language System (UMLS)~\cite{bodenreider2004unified}. 
The resulting knowledge graph contains 11,454 disease entities and 139,143 associated attributes, including 14,303 definitions, 15,938 hypernym relationships, and 108,902 synonyms.

\noindent\textbf{Existing Medical knowledge Databases.}
Disease Ontology (DO) was developed to standardize disease nomenclature and classification, DO offers: \textbf{(i)} disease classification, that encompasses categories such as infectious diseases, genetic disorders, cancers, and metabolic conditions; \textbf{(ii) }disease relationships, that links the diseases as subtypes to their parent categories with a hierarchical structure;
\textbf{(iii)} database mapping, that extensively maps its terms to other medical vocabularies in the Unified Medical Language System (UMLS). UMLS is a comprehensive medical language system crafted by the U.S. National Library of Medicine, that integrates diverse medical terminologies from sources like ICD, SNOMED, and MeSH. It includes: \textbf{(i)} metathesaurus, a compilation of medical concepts that encompass diseases, symptoms, treatments, and more; \textbf{(ii) }semantic network, a framework that captures semantic relationships among medical concepts, detailing connections between diseases and their symptoms, treatments, and causes.

\noindent\textbf{Constructing the Disease Knowledge Graph.}
Starting with disease entities and their attributes from DO, 
we enriched each entity with additional attributes from UMLS through cross-mapping. We then established hypernym relationships from DO as the edges connecting these entities. In total, the knowledge graph encompasses 11,454 disease entities and 139,143 disease attributes, including 14,303 definitions, 15,938 hypernym relationships, and 108,902 synonyms. 
Additionally, for each disease entity, we construct a hierarchical disease chain by linking each disease entity to the root through a random hypernym relation path, as shown in Figure~\ref{fig:methods}A. In the disease chain, the name of each disease entity is randomly chosen from its set of synonyms, further enriching the diversity of the disease representations. This process creates a multi-level hierarchy, where each disease entity is connected to its upper-level relations. The resulting hierarchical structure provides a more nuanced understanding of disease relationships, enhancing the contextualization of each entity and enabling better integration of related knowledge for downstream tasks.

\subsubsection*{Problem Formulation}
Given a set of image-text pairs, denoted by $\mathcal{F} = \{(x_1, c_1), ..., (x_n, c_n)\}$, conventional vision-language pre-training approaches typically employ simple contrastive learning, that aims to align the embeddings of paired images and texts, while simultaneously separating those of unpaired samples:
\begin{equation}
    \text{sim}(\Phi_{\text{v}}(x_i), \Phi_{\text{t}}(c_i)) \gg 
    \text{sim}(\Phi_{\text{v}}(x_i), \Phi_{\text{t}}(c_j)), \text{ } i \neq j,
    \label{eq:421}
\end{equation}
where $\Phi_{\text{v}}$ and $\Phi_{\text{t}}$ denote the visual and the text encoder, respectively. Such training procedure suffers from two issues, 
\textbf{(i)} there is no explicit knowledge injection to the training procedure, for example, the intricate relationships between diseases; \textbf{(ii)} the noise in existing datasets, for instance, low-quality captions and non-pathology images, further introduces ambiguities and inconsistencies in the alignment process.

Herein, we leverage the constructed disease knowledge graph to enhance the vision-language training procedure, from three key aspects: 
a strong text encoder via knowledge representation learning, 
knowledge-guided semantic group construction, and knowledge-guided vision-language representation learning. The target of knowledge-enhanced vision-language training can be formulated as:
\begin{equation}
    \min_{p}{\text{sim}(\Phi_{\text{v}}(\tilde{x}_p), \Phi_{\text{k}}(\tilde{c}_i))} \gg 
    \max_{q}{\text{sim}(\Phi_{\text{v}}(\tilde{x}_q), \Phi_{\text{k}}(\tilde{c}_j))}, \text{ } i \neq j,
    \label{eq:422}
\end{equation}
where $\Phi_{\text{k}}$ denotes a BERT-based knowledge encoder, pre-trained on our constructed disease knowledge graph (KG). $(\tilde{x}_1, ..., \tilde{x}_n, \tilde{c}_i)$ represents the $i$-th semantic group consisting of a single refined caption $\tilde{c}_i$ with a varying number of pathology images $(\tilde{x}_1, ..., \tilde{x}_n)$. The specific components for knowledge enhancement are detailed in the following subsections.

\subsubsection*{Disease Knowledge Encoding}

In this section, we describe the procedure for training a language model to construct a knowledge embedding space, in which attributes of the same disease are mapped to similar embeddings. To account for hierarchical relationships between disease entities, we follow the approach outlined in BioCLIP~\cite{stevens2024bioclip}, 
recursively concatenating each disease entity with its hypernyms (parent nodes) to form hierarchical disease chains, as illustrated in Figure~\ref{fig:methods}A and Figure~\ref{fig:supp_disease_chain}.
This enables to capture the relationships between diseases at different levels of granularity. We then apply metric learning to align the embeddings of disease attributes, ensuring that entities with similar properties are closer in the embedding space.

Specifically, given a set of disease entities with hypernym relations, 
we randomly sample parent nodes to construct hierarchical disease chains for each entity. As a result, each disease entity is associated with a varying number of attributes, including synonyms, disease chains, and definitions, 
which can be denoted by $\mathcal{D} = \{(d_1, \mathbf{a}_1), \dots, (d_n, \mathbf{a}_n)\}$, where $d_i$ denotes the $i$-th disease entity, 
and $\mathbf{a}_i = \{a_i^1, \dots, a_i^k\}$ refer to the associated $k$ attributes, both disease and attributes are in the form of natural language. Note that, for different disease entities, $k$ also varies, our goal here is to train a model that satisfies the following condition:
\begin{equation}
\text{sim}(\Phi_{\text{k}}(a_i^p), \Phi_{\text{k}}(a_i^q)) \gg 
\text{sim}(\Phi_{\text{k}}(a_i^p), \Phi_{\text{k}}(a_j^t)), \text{ } i \neq j,
\label{eq:431}
\end{equation}
\begin{equation}
\text{sim}(\Phi_{\text{k}}(a_i^p), \Phi_{\text{k}}(a_i^q)) = \left<\Phi_{\text{k}}(a_i^p), \Phi_{\text{k}}(a_i^q)  \right> 
\label{eq:432}
\end{equation}
where $\Phi_\text{k}(\cdot) $ denotes the knowledge encoder,
 $\left< \cdot \right>$ refers to the cosine similarity,
$a_i^p, a_i^q$ and $a_j^t$ refer to the randomly sampled attributes from the $i,j$-th disease entity. 
Intuitively, the knowledge encoder is optimised to pull together the attributes of the same disease, while pushing apart attributes of different diseases.

 
\vspace{3pt} \noindent \textbf{Metric Loss.} 
At training time, we employ metric learning to construct an embedding space where the intra-class instances are clustered, and inter-class instances are separated.
Specifically, given a mini-batch with $n$ randomly selected diseases, 
each associated with $k$ attributes, 
we denote the normalized embedding for the $p$-th attribute of the $i$-th disease as:
\begin{equation}\label{eq:433}
\mathbf{z}_{p}^i=\frac{\Phi_{\text{k}}(a_i^p)}{\lVert \Phi_{\text{k}}(a_i^p)\rVert},
\end{equation}
where $\Phi_{\text{k}}(a_i^p)$ represents the embedding of the $p$-th attribute of the $i$-th disease, and $\lVert \cdot \rVert$ denotes the L2-norm. 
We adopt the recently proposed AdaSP loss~\cite{zhou2023adaptive}, 
which finds out a max-min positive similarity and then shapes a loss with the maximal negative similarity:
\begin{equation}\label{eq:434}
    \mathcal{L}_{\text{metic}}=\frac{1}{n}\sum_{i=1}^n{\log \left( 1+\exp\left({(S_{i}^{-}-S_{i}^{+})/\tau} \right)\right)},
\end{equation}

where $\tau$ is a temperature parameter. $S_{i}^{+}$ and $S_{i}^{-}$ denote the max-min positive and the maximal negative similarity, which can be computed by the soft version:
\begin{equation}\label{eq:435}
    S_i^{+} = \max_{p}{\min_{q}{\left< \mathbf{z}_p^i, \mathbf{z}_q^i \right> }} \approx \tau \log \left( \sum_{p=1}^k{\frac{1}{\sum_{q=1}^k{\exp \left({-\left<\mathbf{z}_{p}^{i},\mathbf{z}_{q}^{i}\right> / \tau}\right)}}} \right),
\end{equation}

\begin{equation}\label{eq:436}
    S_i^{-} = \max_{j,p,q}{\left< \mathbf{z}_p^i, \mathbf{z}_q^j \right> }\approx \tau \log \left( \sum_{p=1}^k{\sum_{j=1,j\ne i}^n{\sum_{q=1}^k{\exp \left({\left<\mathbf{z}_{p}^{i},\mathbf{z}_{q}^{j}\right>} / \tau\right)}}} \right). 
\end{equation}

\subsubsection*{Knowledge-guided Dataset Structuring}
In this section, drawing upon the constructed knowledge graph, we propose an automated pipeline for cleaning, and reorganising the public noisy image-text pairs. 

\noindent\textbf{Pathology Image Curation.} 
Quilt-1M~\cite{ikezogwo2024quilt} and OpenPath~\cite{huang2023visual} are publicly available pathology image-language datasets, curated by sourcing images from online platforms, for example, educational videos on YouTube~\footnote{https://www.youtube.com/}, or pathology images from Twitter~\footnote{https://x.com/}. Due to the diverse data sources, these datasets inevitably contain a high noise ratio, for example, radiological images and pathological images may co-exist in the same slide. We therefore train a detector to crop the pathological part from each of the images. Specifically, we manually annotate 1,000 images and then fine-tune a well-established detection model, YOLOv8~\footnote{https://docs.ultralytics.com/}~\cite{wang2023yolov7}, on this annotated dataset, which can be obtained from the website~\footnote{https://huggingface.co/datasets/Loie/PathologyImageDetection-ManualAnnotation}. The refined YOLOv8 model is applied to scan the entire dataset, eliminating samples that fail to detect pathological images, while preserving those with clear pathological detections. Through random sampling and manual verification, we confirm that 99.9\% of the reserved samples consist of pure pathological images.

\noindent\textbf{Knowledge-driven Textual Refinement.}
In open-source datasets, many textual captions are derived from instructional videos, where the spoken language is transcribed directly into texts. As a result, these captions often include substantial amounts of irrelevant information that do not pertain to pathological images. We employ the natural language processing (NLP) tool, SpaCy, to extract named entities from the captions and align them with corresponding entities in the Unified Medical Language System (UMLS)~\cite{bodenreider2004unified}. Sentences that do not contain any UMLS entities are excluded from further processing.

Additionally, as the captions in these datasets are mainly collected from platforms such as Twitter and YouTube, which are typically unstructured and lack explicit disease labels. We perform fuzzy matching between the captions and the synonyms of all disease entities in the pre-constructed disease KG, to identify explicit disease labels for each caption.

\noindent\textbf{Semantic Group Construction.}
With the refined images and texts, we first employ the pathology visual encoder UNI~\cite{chen2024uni} to compute image embeddings and group images with high embedding similarity ({\em e.g.}, consecutive frames in video sequences) using a threshold of 0.95. This step ensures that images of high visual similarity are clustered together into semantic groups. Subsequently, within each group, 
we employ a pre-trained visual-language model MI-Zero~\cite{lu2023visual} to compute embeddings for both images and texts. We then compute a similarity matrix, and only the caption exhibiting the highest similarity to all images within the group is retained, ensuring the most representative description is selected. 
Finally, to resolve potential duplicates or redundant captions across groups, 
we calculate the Intersection over Union (IoU) of their token sets. 
Groups with an IoU greater than 0.9 are merged, resulting in the formation of the final refined semantic groups.

\subsubsection*{Knowledge-enhanced Vision-language Training}

In this section, we present our knowledge-enhanced vision-language training framework by introducing semantic-level alignment via metric learning within well-structured semantic groups. The text encoder, pre-trained on the disease KG, ensures that disease attributes are closely aligned in the embedding space.

\noindent\textbf{Semantic Alignment.}
Given a set of well-structured semantic groups, each with a caption and a varying number of images, we prepare a mini-batch for each iteration in the following steps:
(\textbf{i}) we randomly sample $N$ semantic groups, and for each group (denoted by $\{c_i| i\in [1,\dots, N]\}$), we randomly sample $M$ out of $G$ paired images (denoted by $\{x_{ik}| k\in [1, \dots, M]\}$) with replacement, which yields a positive semantic group;
(\textbf{ii}) we take random crops for each sampled image, and resize the image from $512 \times 512$ pixels to $224 \times 224$ pixels;
(\textbf{iii}) the caption of each group is augmented $M$ times, 
with a 50\% probability to randomly drop 40\% of the words each time. 
Further, we randomly sample a template from the template set (in Table S2)
to repharase the matched captions, 
{\em i.e.}, [Template] + disease label/hierarchical disease chain, 
for instance, {\em a histopathology image of skin squamous cell carcinoma/ skin disease, skin cancer, skin carcinoma, skin squamous cell carcinoma}. 
And we randomly sample one of the dropped captions and the paraphrased caption as the semantics for each tile image.

We denote images and augmented captions in $i$-th semantic group as $(x_i^1, ..., x_i^n, c_i^1, ..., c_i^n)$. Correspondingly, the normalized visual embedding of the $k$-th image in $i$-th semantic group as $\mathbf{v}_{ik} = \Phi_{\text{v}}(x_i^k)$, where $\Phi_{\text{v}}$ suggests the visual encoder. 
Similarly, the normalized embedding of $m$-th caption in $i$-th positive group can be denoted by $\mathbf{t}_{im} = \Phi_{\text{k}}(c_i^m)$, where $\Phi_{\text{k}}$ represents the well-trained knowledge encoder. 
The similarities between positives within each semantic group, correspondingly, can be computed by $\{\mathbf{t}_{im}^T\mathbf{v}_{ik}|m,k\in [1,\dots,M]\}$, 
which composes a matrix with the size of $M\times M$, marked by green boxes in Figure~\ref{fig:methods}C. 
The cosine similarities between negatives, {\em e.g.}, the $i$-th and the $j$-th semantic groups, can be computed by $\{\mathbf{t}_{jm}^T\mathbf{v}_{ik}| m,k\in [1,\dots,M],j \ne i\}$, which composes a matrix with the size of $M\times M(N-1)$.
Overall, the similarity scores for a mini-batch shape a $NM\times NM$ matrix with the diagonal blocks suggesting positive semantic groups, shown in Figure~\ref{fig:methods}C. 

With this similarity matrix, we aim to exploit a metric loss to increase the similarity scores in positive groups, while decreasing scores between negatives.
The metric loss can be formulated by:
\begin{equation}\label{eq:451}
    \mathcal{L}_{\text{metric}}=\frac{1}{N}\sum_{i=1}^N{\log \left( 1+\exp \left((S_{i}^{-}-S_{i}^{+})/\tau\right) \right)},
\end{equation}
where $S^+_i$ and $S^-_i$ represent the positive and the negative similarities for the $i$-th group, and $\tau$ denotes the temperature parameter.

\noindent\textbf{Positive Mining.} There are $M\times M$ positive pairs between augmented images and captions for each semantic group. Their positive embedding similarities can be denoted as 
$\{\mathbf{t}_{im}^T\mathbf{v}_{ik}\, | i\in [1, \dots, N] \}, m,k \in [1,\dots, M]$, where $\mathbf{t}_{im}$ and $\mathbf{v}_{ik}$ suggest the normalized embeddings of the $m$-th caption and the $k$-th image in the $i$-th semantic group. The similarity between the $k$-th image in group $i$ and its hardest caption can be computed by the smooth approximation of the minimum function:
\begin{equation}\label{eq:452}
    S_{ik}^{+}=\min_{m}(\mathbf{t}_{im}^T\mathbf{v}_{ik})\approx - \tau \log \left( \sum_{m=1}^M{\exp \left(-\mathbf{t}_{im}^T\mathbf{v}_{ik}/\tau \right)} \right).
\end{equation}
Intuitively, mining hard positives can accelerate the training procedure and promote the performance of image-text alignment, 
while the hardest positives, to a large extent, could be false positives due to data noise. As a result, we choose a moderate hard positive by mining the easiest one among the hard positive candidates in Eq.~\ref{eq:452}:
\begin{equation}\label{eq:453}
    S_{i}^{+}=\max_k(S_{ik}^{+})\approx \tau \log \left( \sum_{k=1}^M{\exp \left(S_{ik}^{+}/\tau \right)} \right).
\end{equation}
Substituting Eq.~\ref{eq:451} into Eq.~\ref{eq:452}, we have:
\begin{equation}\label{eq:454}
    S_{i}^{+} = \max_k\min_{m}(\mathbf{t}_{im}^T\mathbf{v}_{ik}) \approx \tau \log \left( \sum_{k=1}^M{\frac{1}{\sum_{m=1}^M{\exp \left(-\mathbf{t}_{im}^T\mathbf{v}_{ik}/\tau \right)}}} \right).
\end{equation}
In contrast to mining the hardest positive pair, the least-hard positive is also a moderate choice that can not only reduce the risk of false positives caused by outliers but also prevent the training procedure from degenerating into trivial positives.
It is noteworthy that although the mined $S_{i}^{+}$ is the easiest one among hard positives, it is still a hard positive that achieves an elaborate balance between introducing implicit false positives and collapsing to trivial negatives.

\noindent\textbf{False Negative Elimination.} 
To prevent semantic groups with the same disease label or hypernym relations from being misclassified as negatives, we check if one group is reachable from another via their hypernym paths, setting their negative indicator to zero when applicable.
In contrast to positive mining, we perform the hardest mining for negative samples since we have eliminated the false negatives above. Correspondingly, the negative similarity for the $i$-th semantic group can be formulated as:
\begin{equation}\label{eq:455}
    S_{i}^{-} = \max_k\max_{j,m}(\mathbf{t}_{jm}^T\mathbf{v}_{ik}) \approx \tau \log \left( \sum_{k=1}^M{\left(\sum_{j=1,j\ne i}^{N}\mathcal{I}_{ij} {\sum_{m=1}^M{\exp \left(\mathbf{t}_{jm}^T\mathbf{v}_{ik}/\tau \right)}}\right) } \right).
\end{equation}
where $\mathbf{t}_{jm}$ and $\mathbf{v}_{ik}$ suggest the normalized embeddings of the $m$-th caption in the $j$-th group and that of the $k$-th image in the $i$-th group, respectively.
$\mathcal{I}_{ij}$ is a binary indicator that denotes the negative flag between the $i$-th and the $j$-th semantic group.

\subsubsection*{Model Training Details}

\noindent\textbf{Knowledge Encoding.} 
We adopt the architecture of PubMedBERT~\cite{gu2021domain} to conduct knowledge encoding. The embedding dimension is set to 768. The temperature parameter $\tau$ is set to $0.04$ in Eq.~\ref{eq:434}. The batch size is set to 256, including 32 disease entities with 8 instances per entity. We train the knowledge encoder for 100 epochs with a maximum learning rate of $3\times10^{-5}$ on 4 A100 GPUs.
 
\noindent\textbf{Vision-language Pre-training.} 
KEEP consists of a vision encoder based on the backbone of ViT-L-16 and a text encoder based on the architecture of PubMedBERT. We adopt UNI~\cite{chen2024uni} to initialize the image encoder of KEEP and set the size of the input image as $224\times 224$ pixels. Meanwhile, the text encoder of KEEP is initialized by the pre-trained knowledge encoder. The batch size is set to 128, including 32 semantic groups with 4 image-text pairs per group. The temperature in Eq.~\ref{eq:451} is set to 0.04 across all experiments. We conduct the vision-language pre-training for 10 epochs with a maximum learning rate of $1\times10^{-5}$ on 1 A100 GPU.

\subsubsection*{Zero-shot Evaluation on WSIs}
We first evaluate KEEP on whole slide images to segment cancerous cells, detect malignant tumors, and predict subtypes of cancers. For WSI preprocessing, we follow CONCH~\cite{lu2024visual} to divide each whole slide image into $256\times 256$ tiles ($224\times 224$ pixels for segmentation) at $20\times$ magnification, which are then fed to KEEP to predict the class label of each tile in a zero-shot manner. This evaluation process can be conducted on 1 NVIDIA GeForce RTX 4090 GPU.

\noindent\textbf{Tumor-ratio Based Detection.}
In this work, we utilize tumor-ratio prediction to differentiate between cancerous and non-cancerous WSIs. To establish the theoretical validity of this approach, we formally prove that the probability of classifying a WSI as cancerous is positively correlated with the tumor area fraction present within the slide. The detailed theoretical derivation is provided in Supplementary Note 2.

\noindent\textbf{Tumor-ratio Based Subtyping.}
WSI subtyping can be formulated as a problem of set classification, {\em i.e.}, casting $S = \{x_1, x_2, \dots, x_n\}$ to one of the $C$ categories. 
$C$ represents the number of tumor subtypes.
The set denotes the whole slide, consisting of $n$ independent instances~(tiles in our case), 
and we'd like to compute $P(c_i|S)$, the probability of the set $S$ belonging to class $c_i$. The tumor-area ratio essentially resembles a ratio-based approach for estimating the likelihood: 
\begin{align}
P(c_i | S) = \frac{\# \text{tiles classified as } c_i}{n}
\end{align}
In our case, given the following three conditions:
(i) instances~(tiles) are independent and equally informative,
(ii) there is no prior on potential classes,
(iii) there are large numbers of instances in the set, {\em i.e.,} tiles in the WSIs.
According to the law of large numbers, a ratio-based approach can be used to well approximate the true class likelihood.

\noindent\textbf{Unsupervised Prompt Screening.}
We follow CONCH~\cite{lu2024visual} to randomly concatenate one of the 21 templates and cancer synonyms to generate prompt classifiers. As the zero-shot performance can be sensitive to text prompts, we develop an unsupervised prompt screening approach to improve the robustness of performance on downstream tasks. 

Specifically, given a set of prompt classifiers that aim to categorize $N$ tile images into $C$ types, we have $C$ similarities for each tile image by calculating similarity scores between the tile image and the different text prompts in one prompt classifier. 
Ideally, different categories in one prompt classifier should exhibit consistency in similarity ranges and complementary relationships. For example, in cancer detection tasks, if a tile image has a similarity score of 0.7 with the prompt, \textit{A histopathology image of cancerous tissue}, the corresponding normal prompt (\textit{A histopathology image of normal tissue}) is expected to have a similarity score of 0.3 with the same tile image. Additionally, a larger gap between the similarity scores of different classes indicates better discriminative power. 
To evaluate and refine the prompt classifiers, we introduce a screening score that ranks each prompt classifier based on its range consistency and discriminability.
\begin{equation}\label{eq:471}
    R_s = \sum_{i=1}^{N}\left({S_{i}^{\ast} - S_{i}^{\ast\ast} -|S_{i}^{\ast} + S_{i}^{\ast\ast} - 1|}\right)
\end{equation}
where $S_{i}^{\ast}$, $S_{i}^{\ast\ast}$ denote the largest and the second-largest of the similarities between the $i$-th tile image and the category prompts. 
$S_{i}^{\ast} - S_{i}^{\ast\ast}$ measures the discriminability of the prompt classifier, while $|S_{i}^{\ast} + S_{i}^{\ast\ast} - 1|$ (smaller is better) suggests the range consistency. In this paper, we use this score to screen the top-50 prompt classifiers to evaluate the zero-shot performance on downstream tasks. 

\subsubsection*{Few-shot Finetuning and Five-fold Cross-validation}
For the few‑shot setting, we evaluated both cancer detection and cancer subtyping benchmarks by randomly selecting 15 samples per class as the training pool, with the remaining samples reserved for testing.
From this pool, n‑shot experiments were conducted with $n=1,2,4,8$. For each $n$, we repeated the experiment 10 times using different random seeds, each time sampling $n$ training examples per class from the training pool.
The average performance across the 10 repetitions was reported as the final result.For the five‑fold cross‑validation setting, samples in each class were randomly divided into five equal folds.
In each round, four folds were used for training and the remaining fold for testing, ensuring that all samples were eventually tested once.The reported performance corresponds to the average result across the five folds.

\subsubsection*{Evaluation Metrics}
\noindent\textbf{Performance on WSI Tasks.}
We adopt the same metric as MI-Zero~\cite{lu2023visual} and CONCH, namely, balanced accuracy and weighted F1 to measure the cancer subtyping performance on WSIs. We follow CONCH to use the nonparametric bootstrapping with 1000 samples to construct 95\% confidence intervals for model performance. For cancer segmentation tasks, we exploit area under the curve (AUROC, AUPRC), DICE, and ASSD to evaluate different models. For cancer detection, we use the same metric AUROC, AUPRC, sensitivity, and specificity as CHIEF~\cite{wang2024pathology} to evaluate the performance of different models.

\noindent\textbf{Zero-shot Performance on Tile-level Tasks.}
For classification tasks, we adopt the same metric as PLIP~\cite{huang2023visual}, namely, weighted F1 (wF1). 
For cross-modal retrieval tasks, we adopt the metric of Recall@K, suggesting the ratio of correctly retrieved queries in Top-K retrieved samples.
We also use the nonparametric bootstrapping with 1000 samples to construct 95\% confidence intervals for tile-level tasks.

\subsection*{Quantification and statistical analysis}

For cancer region segmentation tasks, paired t-test is used to assess the statistical significance between the performance distributions of different models. n.s. denotes not significant, ** denotes $P < 0.01$, and *** denotes $P < 0.001$. We use the nonparametric bootstrapping with 1000 samples to construct 95\% confidence intervals for tile-level tasks. Two-sample t-test is used to assess the
statistical significance of predicted tumor ratios among different WSIs. In all boxplot visualizations, the box elements represent the first quartile (25th percentile), median (50th percentile), and third quartile (75th percentile) values, while the whiskers encompass data points falling within 1.5 times the interquartile range (IQR) from either quartile.

\clearpage
\section*{Supplementary}

\renewcommand{\thefigure}{S\arabic{figure}}
\renewcommand{\figurename}{Figure}
\renewcommand{\thetable}{S\arabic{table}}
\renewcommand{\tablename}{Table}
\setcounter{figure}{0}
\setcounter{table}{0}

\noindent \textbf{Supplementary Note 1. Ablation Studies. }

~\\
\textbf{Ablations on knowledge enhancement.}
To validate the enhancement of disease knowledge, we compare the performance of KEEP (knowledge-enhanced) with that of na\"ive contrastive baseline (same backbone, without knowledge enhancement) on all WSI tasks, The experimental results are shown in Figure 5E, Figure~\ref{fig:supp_sub}C-D, and Table S5.

\noindent \textbf{Ablations on tumor-ratio.}
We also conducted an ablation study on the tumor‑ratio aggregation strategy, using the top‑100 pooling method from CONCH as a comparison. As shown in Table S5,
KEEP‑Top100 and KEEP‑Ratio represent evaluations of the same model, differing only in the aggregation strategy used during WSI‑level inference when deriving the final slide label.

\noindent \textbf{Ablations on semantic groups.}
We conduct the ablation on semantic groups, where the baseline adopts image-text pairs to perform contrastive learning
with the same backbone as KEEP. The experimental results on text-to-image and image-to-text retrieval tasks are shown in Table S5.

\noindent \textbf{Ablations on caption augmentation.}
We perform an ablation study to assess the effect of caption augmentation.
The baseline model follows the exact same training procedure as KEEP, differing only in the absence of caption augmentation during vision–language alignment pretraining. The experimental results are shown in Table S5.

\noindent \textbf{Ablations on loss function.}
We perform an ablation study on the loss function of knowledge encoding, examining different combinations of positive and negative sampling strategies.
Specifically, we evaluate four variants:
a. hardest positive \& hardest negative,
b. least‑hard positive \&least‑hard negative,
c. hardest positive \& least‑hard negative, and
d. the configuration adopted by KEEP, which employs the least‑hard positive \& hardest negative pairs.
The experimental results are shown in Table S5.

\noindent \textbf{Ablations on knowledge graph.}
To assess the impact of knowledge graph construction, we conducted two ablation studies:
\textbf{a. Removing non-cancer diseases.}
We re-trained the text encoder with non-cancer diseases removed. The experimental results of downstream WSI-level benchmarks are shown in Figure~\ref{fig:supp_sub}F.
It can be seen that cancer-only nodes show an average of 1.1\%  and 2.5\% decline compared to the full ontology in cancer detection and subtyping benchmarks, suggesting that the broader disease categories may still contribute to generalization in cancer-related tasks.
\textbf{b. Removing disease relationships.}
We also performed pre-training and vision–language alignment without incorporating disease relationships. The experimental results, shown in Figure~\ref{fig:supp_sub}G,
reveal a 3.4\% performance decrease in cancer subtyping benchmarks, whereas no performance degradation is observed in cancer detection tasks. This likely occurs because samples within the same disease branch cannot be excluded from the negative pairs under this setting, highlighting that inter-disease relationships play an important role in fine-grained cancer subtyping.

\noindent \textbf{Ablations on text encoder.}
To assess whether comparable performance could be obtained by directly leveraging embeddings from large pre-trained language models. We conduct three ablation studies with different text encoders:
\textbf{a. BERT Without knowledge pre-training}. 
We align the vision encoder with PubMedBERT (pre-trained on all PubMed abstracts and full texts) without any additional knowledge-driven pre-training. As shown in Figure~\ref{fig:supp_sub}H-I,
the performance was significantly lower compared to our knowledge-enhanced approach, highlighting the benefit of incorporating structured medical ontologies.
\textbf{b. Clinical-Longformer without knowledge pre-training}.
We also employed Clinical-Longformer, a transformer model pre-trained on large-scale clinical notes---as the text encoder in our vision-language alignment training. Similar to the PubMedBERT setting, no knowledge-driven pre-training was applied. We then comprehensively evaluated this model on all downstream tasks. As shown in Figure~\ref{fig:supp_sub}H-I,
clinical-longformer achieves a comparable performance to that of the BERT-based model without knowledge enhancement. which is consistently lower than that of our knowledge-enhanced framework, demonstrating that structured medical knowledge complements domain-specific language models and substantially improves fine-grained medical understanding.
\textbf{c. LLM baselines}. 
We also extracted text embeddings from OpenAI LLM (named \textbf{``text-embedding-3-large''}, pre-trained on a large-scale clinical corpus) for all captions, and aligned them with vision features. We then evaluated this setup on the downstream tasks. The results are illustrated in Figure~\ref{fig:supp_sub}H-I.
As expected, this configuration yielded the lowest performance among all tested settings, as the language model are optimized for broad semantic coverage rather than fine-grained biomedical discrimination, lack explicit grounding to the disease ontology or image-specific context. Consequently, such general-purpose representations may fail to capture subtle subtype-related visual-textual correspondences required for precise cancer subtyping.

\clearpage

\noindent\textbf{Supplementary Note 2}

\begin{proof}
The proof establishes monotonic dependence through these steps:
\begin{theorem}[\textbf{Positive Correlation of Tumor-ratio and Cancer Probability}]
Let $A_c(p)$ denote the proportion of tumor region in a pathological patch $p$, and $P(C|p)$ denote the probability that a classifier predicts $p$ as cancerous. Under a linear feature representation and sigmoid classifier with significant tumor feature contribution, $P(C|p)$ is positively correlated with $A_c(p)$.
\end{theorem}

\begin{assumption}[Linear Feature Representation]
The feature map $F(p)$ combines tumor ($f_c$) and non-tumor ($f_n$) features linearly:
\begin{equation}
F(p) = A_c(p) \cdot f_c + (1 - A_c(p)) \cdot f_n
\label{eq:feature}
\end{equation}
where $f_c, f_n \in \mathbb{R}^d$ are feature vectors representing unit tumor and non-tumor regions.
\end{assumption}

\begin{assumption}[Sigmoid Classifier]
The cancer probability uses a sigmoid activation:
\begin{equation}
P(C|p) = \sigma\big(w \cdot F(p) + b\big) = \frac{1}{1 + e^{-(w \cdot F(p) + b)}}
\label{eq:classifier}
\end{equation}
where $w \in \mathbb{R}^d$ is the weight vector and $b \in \mathbb{R}$ is the bias.
\end{assumption}

\begin{assumption}[Tumor Feature Significance]
The net tumor contribution dominates:
\begin{equation}
k = w \cdot (f_c - f_n) > 0
\label{eq:significance}
\end{equation}
This captures the clinically relevant property that tumor features contribute more strongly to cancer detection than non-tumor features.
\end{assumption}

Substitute \eqref{eq:feature} into \eqref{eq:classifier}:
\begin{align}
P(C|p) &= \sigma\Big(w \cdot \big[A_c(p) f_c + (1 - A_c(p)) f_n\big] + b\Big) \nonumber \\
&= \sigma\Big(A_c(p) w \cdot (f_c - f_n) + w \cdot f_n + b\Big) \label{eq:expanded}
\end{align}

Define $c = w \cdot f_n + b$ and apply \eqref{eq:significance}:
\begin{equation}
P(C|p) = \sigma\big(k \cdot A_c(p) + c\big) \quad \text{with} \quad k > 0
\label{eq:simplified}
\end{equation}

Since $\sigma(x) = (1 + e^{-x})^{-1}$ is strictly increasing $\forall x \in \mathbb{R}$, and $k > 0$:
\begin{equation}
\frac{\partial P(C|p)}{\partial A_c(p)} = \sigma'\big(k \cdot A_c(p) + c\big) \cdot k > 0 \quad \forall A_c(p) \in [0,1]
\label{eq:derivative}
\end{equation}
where $\sigma'(x) = \sigma(x)(1 - \sigma(x)) > 0$ for all finite $x$.

Thus $P(C|p)$ is strictly increasing in $A_c(p)$: For any patches $p_1$, $p_2$,
\begin{equation}
A_c(p_1) > A_c(p_2) \implies k \cdot A_c(p_1) + c > k \cdot A_c(p_2) + c \implies P(C|p_1) > P(C|p_2)
\end{equation}
proving positive correlation. \qedhere
\end{proof}

\clearpage

\begin{figure}[!t]
    \centering
    \includegraphics[width=\textwidth]{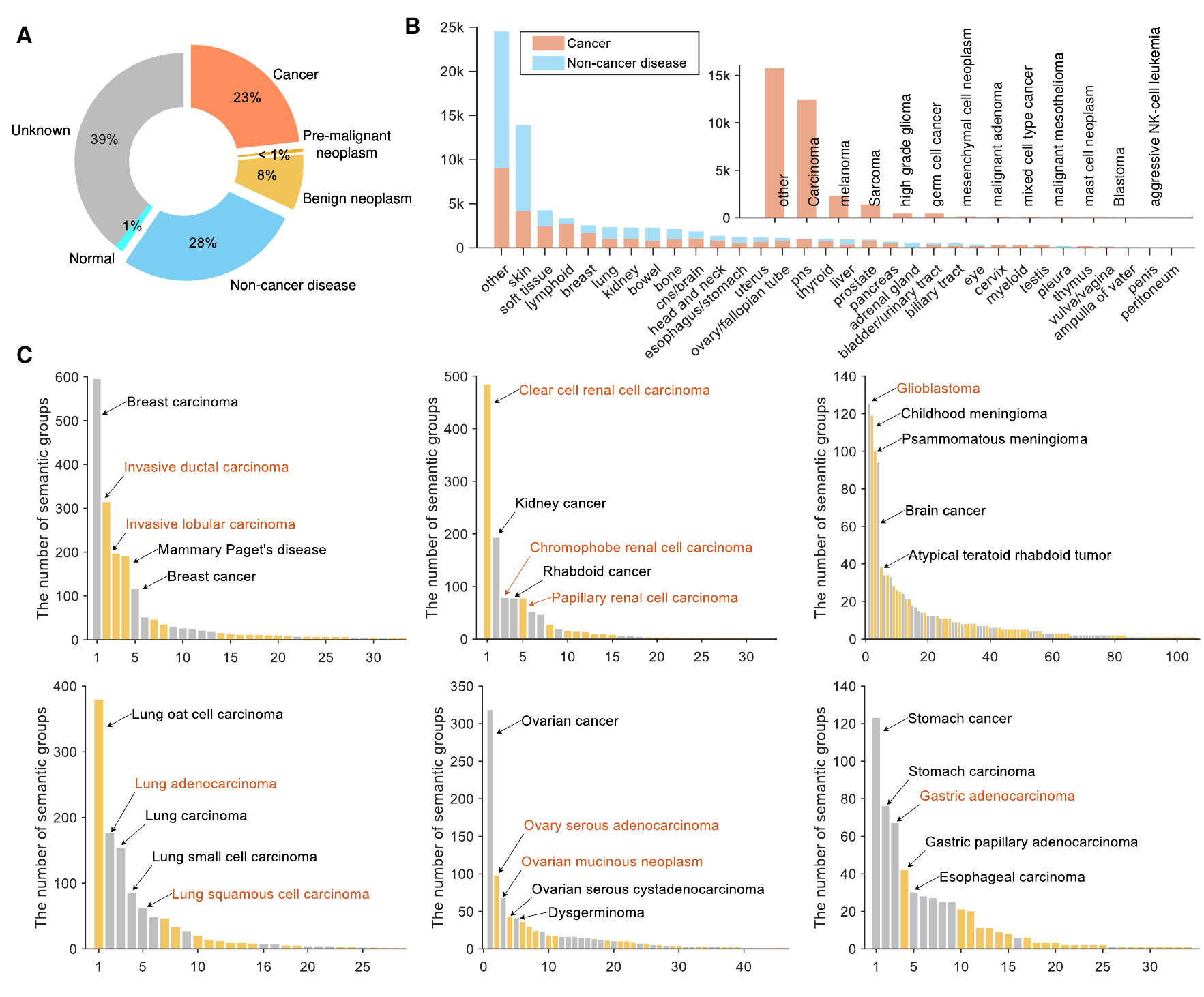}
    \caption{\textbf{Statistics of semantic groups.} Related to Figure 1.
    \textbf{(A)} Statistics of all semantic groups, organized by structuring one million noisy pathology image-text pairs with the guidance of disease KG. More than 60\% semantic groups are linked to specific disease nodes. 
    \textbf{(B)} The anatomy and cell type distribution of the semantic groups with known disease labels. The term "other" in the larger/smaller bar graph suggests that the anatomy/the tumor cell type is unspecified or unknown. The anatomical taxonomy is based on OncoTree~\cite{kundra2021oncotree}. The anatomy and tumor types with the largest number of semantic groups are skin and carcinoma, respectively.
    \textbf{(C)} Distributions of cancer types, for example, anatomies, including breast, kidney, brain/CNS, lung, ovarian, and esophagus/stomach. The top 5 cancer subtypes are listed by the arrowed text. The gray and yellow bars represent the non-leaf and leaf nodes in the constructed knowledge base. The red text suggests the subtypes overlapped with the cancer subtyping tasks in this paper.
    }
    \label{fig:supp_statictis}
\end{figure}

\begin{figure}
    \centering
    \includegraphics[width=0.95\linewidth]{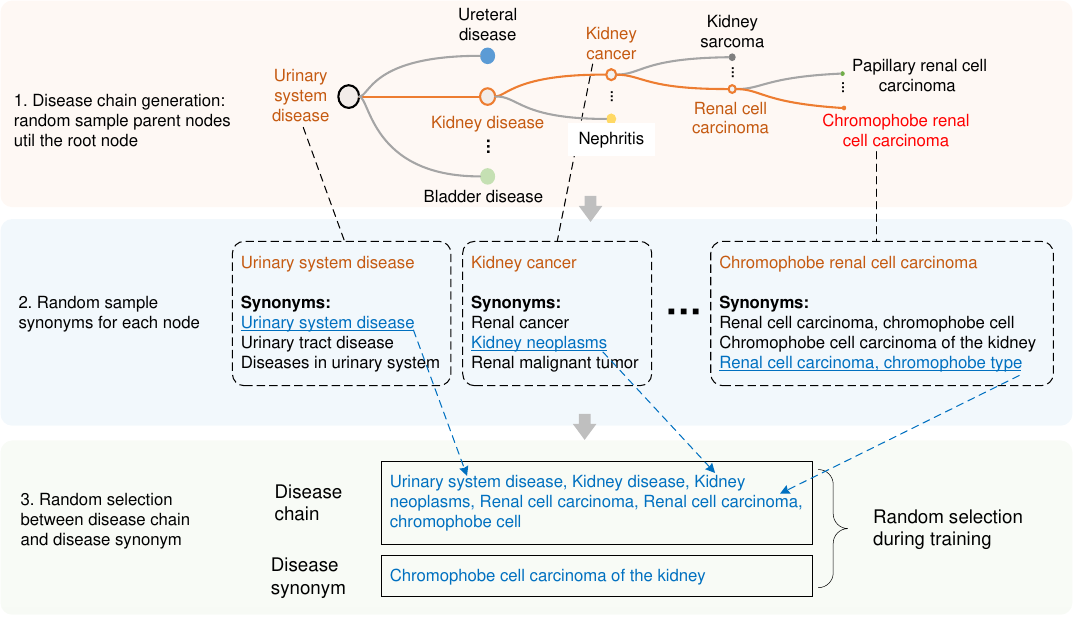}
    \caption{\textbf{The pipeline of disease chain generation}. Related to Figure 2.  For "\textbf{Chromophobe renal cell carcinoma}" during knowledge encoding: (i) Our knowledge encoding constructs a disease chain for each disease node, extending from the current concept to the root node, as shown in;
    (ii) Each node in this hierarchical chain is represented by randomly sampled synonyms from our curated knowledge graph;
    (iii) During training, we align the text embedding of the current disease concept (randomly sampled synonym) with that of its definition, as well as with that of the corresponding disease chain, to learn hierarchical relationships among disease ontology. }
    \label{fig:supp_disease_chain}
\end{figure}

\begin{figure}[!t]
    \centering
    \includegraphics[width=\textwidth]{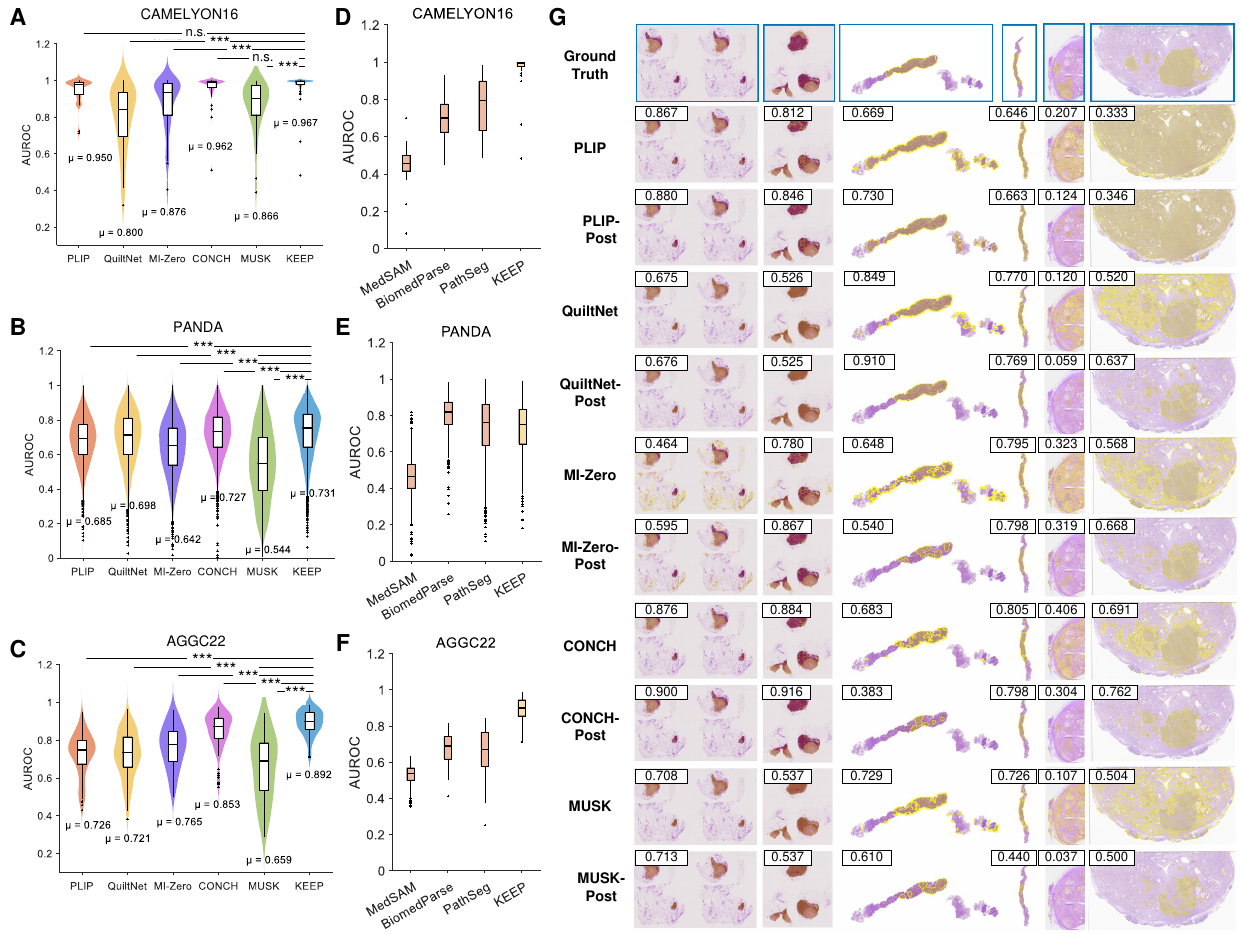}
    \caption{\textbf{AUROC performance of cancer region segmentation on three datasets and exemplary visualization results of different models.} Related to Figure 3. 
    \textbf{(A-C)} Performance comparisons of AUROC for various models, including PLIP, QuiltNet, MI-Zero, CONCH, and MUSK, and our proposed KEEP, across three WSI datasets: CAMELYON16 (48 WSIs), PANDA (10,494 WSIs), and AGGC22 (128 WSIs), with $\mu$ indicating the average performance. The paired \textit{t} test is used to assess the statistical significance between the performance distributions of different models. n.s. represents non-significant, $**$ denotes $P\ <\ 0.01$, and $***$ denotes $P\ <\ 0.001$.
    \textbf{(D-F)} Performance comparisons of AUPRC between KEEP and text-based segmentation models, including MedSAM, BiomedParse, and PathSeg. The box plots present the median, first, and third quartiles of results.
    \textbf{(G)} Exemplary WSIs from three datasets (the first two for CAMELYON16, the middle two for PANDA, and the last two for AGGC22) showing ground truth and predicted segmentation masks of different models. The number in the top-left of each result image suggests the DICE score.
    }
    \label{fig:supp_seg}
\end{figure}

\begin{figure}[!t]
    \centering
    \includegraphics[scale=0.8]{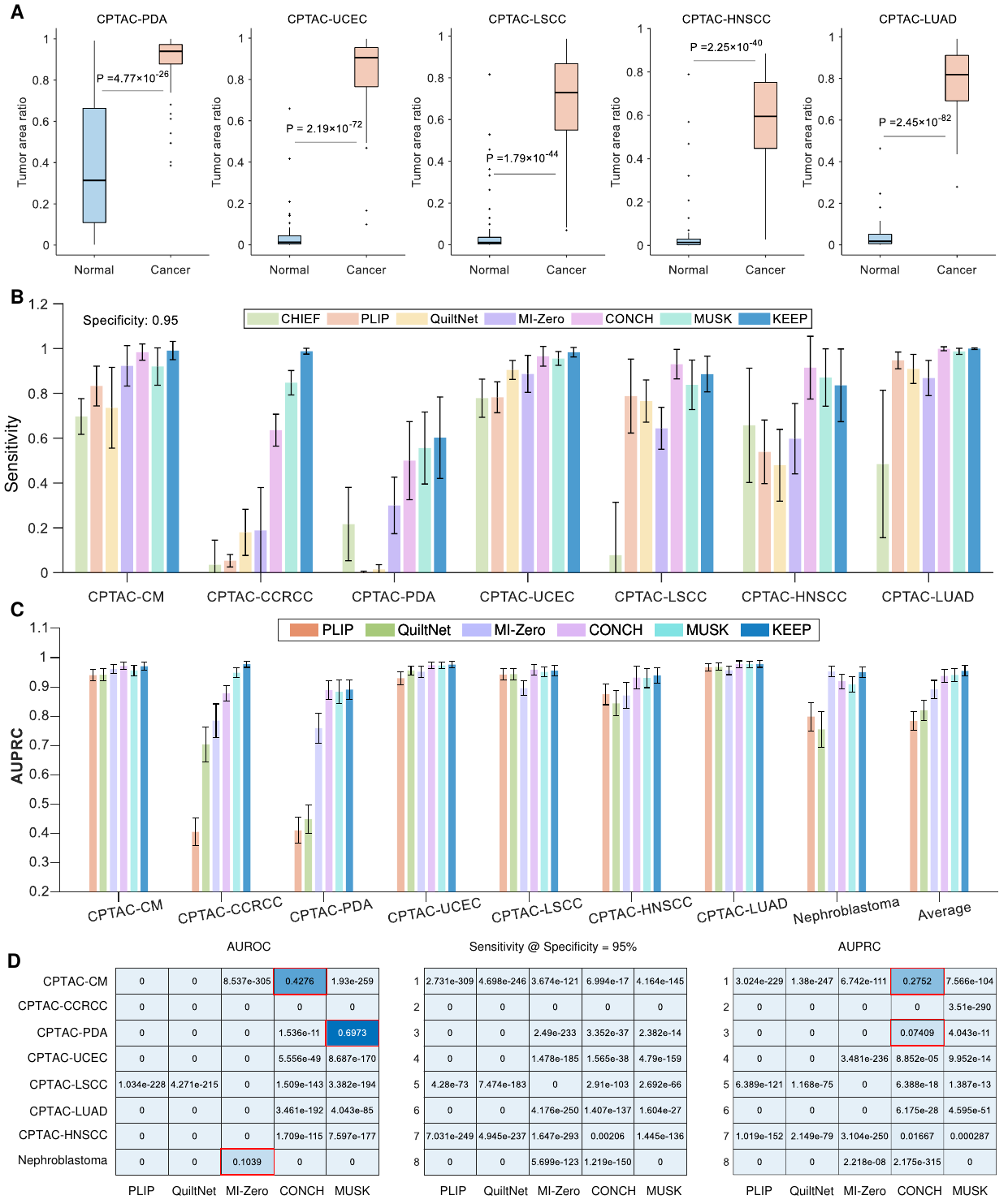}
    \caption{\textbf{Additional results on zero-shot cancer detection.} Related to Figure 4.
    \textbf{(A)} The comparison of the predicted tumor ratio between normal and cancer WSIs
    \textbf{(B)} Comparison of cancer detection sensitivities on each dataset at the specificity of 0.95, the error bar denotes the standard deviation of the performance with 1,000 bootstrap iterations.
    \textbf{(C)} The average AUPRC performance of different models across all cancer detection benchmarks. The error bar denotes the standard deviation of the performance.
    \textbf{(D)} P-values between KEEP and other foundation models across all cancer detection datasets. Red boxes suggest that $P > 0.05$.
    }
    \label{fig:supp_det}
\end{figure}

\begin{figure}[!t]
    \centering
    \includegraphics[scale=0.73]{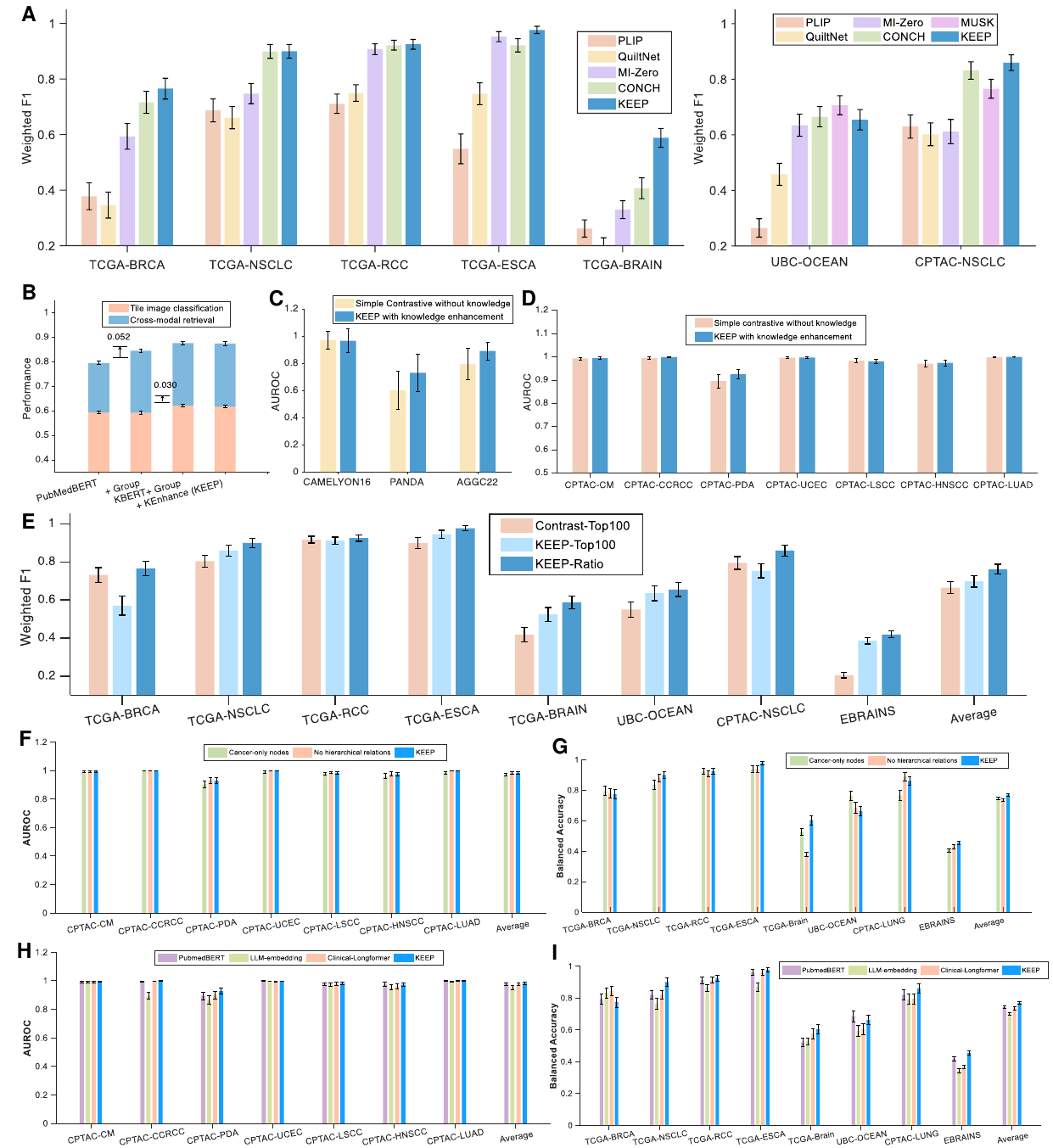}
    \caption{\textbf{Additional results on zero-shot cancer subtyping and ablation study.} Related to Figure 5.
    \textbf{(A)} Comparison of weighted F1 across different models on seven datasets with common cancer subtypes.
    \textbf{(B)} Comparison of performance across different experimental settings. ”PubMedBERT” indicates that the text encoder is initialized using PubMedBERT, with naive contrastive learning applied to align images and their paired captions. ”+ Group” signifies that semantic group alignment is employed instead of direct image-text pair alignment. ”KBERT + Group” incorporates a knowledge encoder alongside semantic group alignment. ”+ KEnhance (KEEP)” further builds on ”KBERT + Group” by introducing knowledge-enhanced caption augmentation and strategies for eliminating false negatives.
    \textbf{(C-D)} Performance comparison between naive contrastive without knowledge and KEEP with knowledge enhancement on the task of zero-shot cancer region segmentation and cancer detection, respectively. 
    \textbf{(E)} Performance comparison of weighted F1 between naive contrastive with Top-100 pooling strategy (Contrast-Top100), KEEP with Top-100 pooling strategy (KEEP-Top100) and KEEP with tumor-ratio strategy (KEEP-Ratio).
    Average performance comparison of different knowledge ablation studies for cancer detection (\textbf{F}) and cancer subtyping (\textbf{G}). Average performance comparison of different text encoder ablations for cancer detection (\textbf{H}) and cancer subtyping (\textbf{I}). The error bar denotes the standard deviation of the performance.
    }
    \label{fig:supp_sub}
\end{figure}

\begin{figure}[!h]
    \centering
    \includegraphics[scale = 0.6]{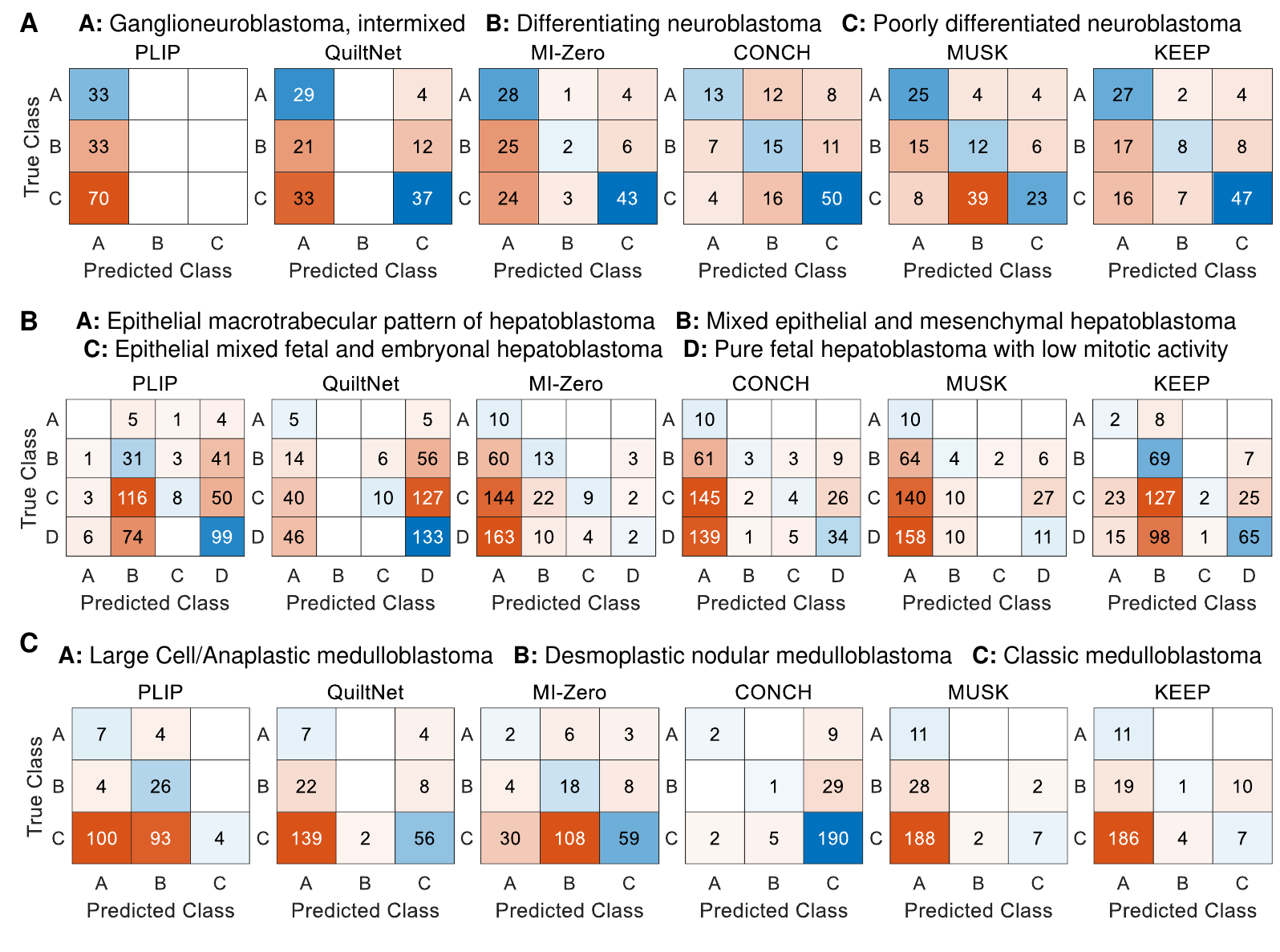}
    \caption{\textbf{Confusion matrices of zero-shot performance for different models on different datasets}. Related to Figure 6. \textbf{(A)} Neuroblastoma. \textbf{(B)} Hepatoblastoma. \textbf{(C)} Medulloblastoma.}
    \label{fig:supp_xinhua}
\end{figure}

\end{document}